\documentclass[sigconf]{acmart}
\usepackage{multicol}
\usepackage{multirow}

\usepackage{tikz}

\usepackage{stmaryrd}

\usepackage{amsmath,amsfonts}
\usepackage{algorithmic}
\usepackage{float}
\usepackage{array}
\usepackage[caption=false,font=footnotesize]{subfig}
\usepackage{textcomp}
\usepackage{stfloats}
\usepackage{url}
\usepackage{verbatim}

\DeclareMathAlphabet{\mathpzc}{OT1}{pzc}{m}{it}
\usepackage{float}
\usepackage{hyperref}
\usepackage{xurl}

\usepackage{algorithmic}
\usepackage[linesnumbered,ruled,vlined]{algorithm2e}
\usepackage{enumitem}
\usepackage{mathtools}

\usepackage{textcomp}
\usepackage{subfig}
\usepackage{soul}
\usepackage{orcidlink}

\usepackage{mdframed,lipsum,calc}
\newtheorem{proof*}{Proof}

\usepackage{balance}

\newcommand{\FIRST}{FIRST}
\newcommand{\first}{FIRST}

\usepackage{mathrsfs}
\usepackage{upgreek} 
\usepackage{graphicx,xcolor,amsmath,amsfonts,verbatim}

\def\BibTeX{{\rm B\kern-.05em{\sc i\kern-.025em b}\kern-.08em
    T\kern-.1667em\lower.7ex\hbox{E}\kern-.125emX}}

\newcommand{\change}[1]{\textcolor{black}{#1}}
\newcommand{\revision}[1]{\textcolor{black}{#1}}

\newcommand{\set}[1]{ \mathbb {#1} }
\newcommand{\coordinator}{ \mathcal {C} }
\newcommand{\messages}{ \emph {M} }
\newcommand{\verifiers}{\mathbb {V} }
\newcommand{\verifpr}{\mathbb {V}^{\prime} }
\newcommand{\VDF}{\mathcal {V} }
\newcommand{\sigagg}{\sigma_{\mathsf{agg}}}
\newcommand{\sigaggpr}{\sigma^{\prime}_{\mathsf{agg}}}
\newcommand{\bc}{\mathit{BC}}

\newcommand{\dac}{\emph{dAC}}
\newcommand{\Tau}{T}
\newcommand{\F}[1]{\mathcal{F}_{\text{#1}}}
\newcommand{\BC}{\mathit{BC}}

\newcommand{\firsttip}{\mathit{\first\_FEE}}
\newcommand{\firstfee}{\mathit{\first\_FEE}}

\newcommand{\sign}{\mathsf{Sign}}
\newcommand{\KeyGen}{\mathsf{KeyGen}}
\newcommand{\verify}{\mathsf{Verify}}
\newcommand{\agg}{\mathsf{agg}}

\newcommand{\Agg}{\mathsf{Agg}}
\newcommand{\aggsign}{\mathsf{Aggregate}}
\newcommand{\aggverify}{\mathsf{AggregateVerification}}
\newcommand{\eval}{\mathsf{Eval}}
\newcommand{\setup}{\mathsf{Setup}}

\newtheorem{theorem}{Theorem}[section]
\newtheorem{definition}{Definition}[section]

\newcommand{\deploy}{{\mathsf{deploy}}}
\newcommand{\invoke}{{\mathsf{invoke}}}
\newcommand{\getData}{{\mathsf{getData}}}
\newcommand{\getBlockNum}{{\mathsf{getBlockNum}}}
\newcommand{\verkey}{\mathsf{Verification Key}}
\newcommand{\Signature}{\mathsf{Signature}}
\newcommand{\Verified}{\mathsf{Verified}}
\newcommand{\ctable}{\mathrm{cTable}}
\newcommand{\atable}{\mathrm{aTable}}
\newcommand{\stable}{\mathrm{sTable}}
\newcommand{\utable}{\mathrm{uTable}}
\newcommand{\idtable}{\mathrm{idTable}}
\newcommand{\bctable}{\mathrm{bcTable}}
\newcommand{\bcblockmaxfee}{\mathrm{blockMaxFee}}
\newcommand{\blockHashTime}{\mathrm{bcHashTime}}
\newcommand{\hashtable}{\mathrm{hashTable}}

\newcommand{\ctrTime}{\mathrm{ctrTime}}
\newcommand{\blocknum}{\mathrm{blockNum}}
\newcommand{\txpooltable}{\mathrm{txpoolTable}}
\newcommand{\sctable}{\mathrm{scTable}}
\newcommand{\adv}{\mathcal{A}}

\newcommand{\simu}{\mathcal{S}}
\newcommand{\env}{\mathcal{Z}}

\usepackage{tabularx}
\usepackage[
n ,
advantage ,
operators,
sets,
adversary ,
landau ,
probability ,
notions, 
logic,
ff,
mm,
primitives,
events,
complexity,
asymptotics ,
keys
]{cryptocode}

\begin{document}

\title{\texorpdfstring{\FIRST: \underline{F}rontrunn\underline{I}ng \underline{R}esistant \underline{S}mart Con\underline{T}racts}
                     {FIRST: Frontrunning Resistant Smart Contracts}}
\pagestyle{plain}




\author{Emrah Sariboz, Gaurav Panwar, Roopa Vishwanathan, Satyajayant Misra}
\email{{emrah, gpanwar, roopav, misra}@nmsu.edu}
\affiliation{%
  \institution{New Mexico State University}
  \country{Las Cruces, NM, USA}
}

\begin{abstract}
Owing to the increasing acceptance of cryptocurrencies, there has been widespread adoption of traditional financial applications such as lending, borrowing, margin trading, and more, into the cryptocurrency realm. In some cases, the inherently transparent and unregulated nature of cryptocurrencies exposes users of these applications to attacks. One such attack is \emph{frontrunning}, where a malicious entity leverages the knowledge of currently unprocessed financial transactions and attempts to get its own transaction(s) executed ahead of the unprocessed ones. The consequences of this can be financial loss, inaccurate transactions, and even exposure to more attacks. We propose \FIRST{}, a framework that prevents frontrunning, and as a secondary effect, also backrunning and sandwich attacks. \FIRST{} is built using cryptographic protocols including verifiable delay functions and aggregate signatures. We formally prove the security of \FIRST{} using the universal composability framework, and experimentally demonstrate its effectiveness using Ethereum and Binance Smart Chain blockchain data. We show that with \FIRST{}, the probability of frontrunning is approximately 0.00004 (or 0.004\%) on Ethereum and 0\% on Binance Smart Chain, making it effectively near zero.
\end{abstract}

\keywords{Frontrunning Prevention, Smart Contract Security, MEV Mitigation, Transaction Ordering Fairness.}

\maketitle 
\section{Introduction} 
\label{introduction}
The decentralized, trustless, and censor-resistant nature of Ethereum, along with its support for smart contracts, has enabled a wide range of financial applications and has created the Decentralized Finance (DeFi) ecosystem, which is worth more than 75 billion USD as of December 2024~\cite{DefiLama}. With the recent developments, many real-world financial products such as money lending and borrowing, margin trading, exchange platforms, derivatives and more, are being made available to the blockchain users via smart contracts~\cite{compound,dydx,UNISWAP}. Unfortunately, the absence of regulations allows malicious actors to adopt and employ dubious practices from traditional finance within the cryptocurrency ecosystem. 

\begin{figure}[t]
\includegraphics[width=0.5\textwidth]{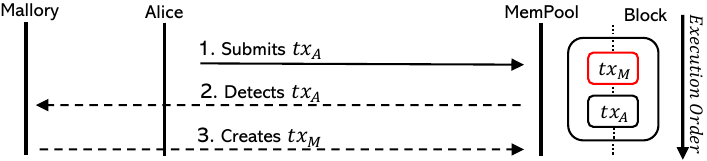}
\caption{Steps involved in a frontrunning attack.}
\label{tbl:sandwich1}
\centering
\vspace{-2em}
\end{figure}

In finance, frontrunning is an act of purchasing stock or other securities right before a large (whale) transaction owing to access to non-public information. By doing so, one can take advantage of the outcomes of large unprocessed transactions to be executed after a later time than one's own. Frontrunning has been classified as illegal by monitoring entities, such as the U.S. Securities and Exchange Commission (SEC) and principally prevented by extensive regulations~\cite{SEC}. In permissionless chains such as Ethereum, transactions that do not utilize private relayers like Flashbots~\cite{flashbots} or directly interact with validators to obscure their details are publicly visible in the pending pool (or mempool) before being processed. This visibility allows adversaries, such as Mallory, to monitor the peer-to-peer (P2P) network for potentially exploitable transactions. For example, when an honest user, Alice, submits a transaction $tx_{A}$ with a gas price of $G_{A}$, Mallory can craft a competing transaction $tx_{M}$ with a higher gas price $G_{M}$, where $G_{M} > G_{A}$, ensuring $tx_{M}$ is prioritized and included in the next block before $tx_{A}$. Figure~\ref{tbl:sandwich1} illustrates a typical frontrunning attack.

Examples of frontrunning attacks can be seen on various decentralized applications (dApps).
The first and most prominent attack vector is on decentralized exchanges (DEXes). DEXes are exchange platforms built on smart contracts that enable users to exchange assets without the need for an intermediary~\cite{qin2021quantifying}.
Unlike centralized exchanges, where users wait for their buy/sell orders to be fulfilled, most DEXes—e.g., Uniswap~\cite{UNISWAP}—use an automatic pricing mechanism known as an Automated Market Maker (AMM) to perform instant trades.
%
A frontrunner can perform attacks with highly predictable results due to deterministic pricing mechanism as well as the transparency of liquidity amounts of decentralized exchanges. In this context, Qin \emph{et al.} estimated a profit of 1.51 Million USD made by frontrunners~\cite{qin2021quantifying}. Other domains that are affected by frontrunning attacks include (but are not limited to) gambling \cite{eskandari2019sok}, bug bounty programs \cite{breidenbach2018enter}, smart contract exploits \cite{torres2021frontrunner}, and clogging \cite{qin2021quantifying}, which emphasizes the threat and the need for mitigation.
In this work, we aim to mitigate frontrunning attacks on blockchains that support smart contracts such as Ethereum, but without modifying the blockchain's underlying infrastructure. 
The core idea behind \FIRST{} is to prevent Mallory from frontrunning Alice's transaction, $tx_{A}$, with her own transaction, $tx_{M}$, by ensuring that $tx_{A}$ is included in the block before $tx_{M}$. We introduce a novel approach to achieve this by leveraging Verifiable Delay Functions (VDFs) to impose a delay on $tx_{M}$~\cite{boneh2018verifiable}. Specifically, \FIRST{} requires Mallory's transaction to wait for a predetermined amount of time, during which the VDF is evaluated, before it can interact with the dApp implementing the frontrunning protection. Once Mallory completes the VDF evaluation, she generates a proof that is then verified by a set of verifiers. Transactions that fail to be verified by the verifiers are rejected from calling the smart contract function. 

Importantly, \FIRST{} is not limited to a specific application and can be employed by various dApps, such as auctions, decentralized name services, NFT marketplaces, and others that are susceptible to frontrunning attacks. \FIRST{} helps protect users from frontrunning and backrunning attacks, and consequently from sandwich attacks as well~\cite{zhou2020high}. \revision{FIRST operates entirely at the application layer, requiring no changes to the underlying blockchain or its consensus protocol, thus requiring no changes to the blockchain framework. The specific FIRST transactions require an additional amount of verification by the validators (miners), which is incentivized by the framework to ensure the transactions are picked up to be verified and added to the chain.}

Our novel {\bf contributions} are as follows:
{\bf a)} We propose FrontrunnIng Resistant Smart ConTracts (\FIRST{}), a general-purpose solution to the frontrunning problem using cryptographic protocols, such as VDFs and aggregate signatures~\cite{boneh2003aggregate}. \FIRST{} significantly curtails frontrunning attacks in EVM-based blockchains while requiring no changes to the underlying blockchain infrastructure. As an application-agnostic solution, \FIRST{} can be easily adopted by any dApp.  {\bf b)} We discuss the effectiveness of \FIRST{} and experimentally evaluate it using Ethereum and Binance Smart Chain transaction data. {\bf c)} We rigorously prove the security of \FIRST{} using the Universal Composability (UC) framework. 

\noindent \textbf{Paper Organization}: 
In Section~\ref{preliminaries}, we provide a concise explanation of relevant preliminary concepts. Section~\ref{sysmodel} presents the system model and threat model. In Section~\ref{construction}, we detail the construction of \FIRST{} and its constituent protocols. Section~\ref{scenarios} offers a comprehensive security analysis of \FIRST{}. In Section~\ref{implementation}, we elaborate on the implementation and evaluation of our system. We discuss the design choices and limitations of \FIRST{} in Section~\ref{sec:limitation}. We review related literature in Section~\ref{relatedworknew} and conclude the paper in Section~\ref{conclusion}.

\section{Preliminaries}

\label{preliminaries}
\subsection{Ethereum and DeFi}
 

Bitcoin demonstrated blockchain technology's potential by enabling direct transactions between untrusted parties without a central authority. Ethereum expanded on this by introducing \emph{smart contracts}—self-executing programs on the blockchain that activate when predefined conditions are met. Developed using languages such as Solidity or Vyper, these transparent smart contracts have led to the creation of dApps. Finance-related dApps have enabled DeFi, which is an umbrella term that includes various financial products (such as flash loans, asset management services, decentralized derivatives, and insurance services) available to any user with an internet connection in a decentralized manner~\cite{aave,yearn}. 
It allows users to utilize financial products at any time.
Additionally, DeFi products enable end-users to employ them in a non-custodial fashion, giving users complete control over their money, as opposed to traditional financial services based on a custodial model. 

\subsection{Cryptographic Preliminaries}
\label{sec:cryptoprelim}

\noindent \textbf{Verifiable Delay Function}: A Verifiable Delay Function (VDF) is a deterministic function $f \colon X \to Y$ requiring a fixed number of sequential steps, $\Tau$, to compute, with efficient public verification~\cite{boneh2018verifiable}. While time-lock puzzles~\cite{rivest1996time,cryptoeprint5} also enforce $\Tau$ sequential steps, they lack public verifiability and focus on encryption, making them unsuitable for applications like FIRST{}, where proof of elapsed time is essential. Recent VDF constructions by Pietrzak and Wesolowski~\cite{pietrzak2018simple, wesolowski2019efficient} address these limitations. We adopt Wesolowski’s VDF~\cite{wesolowski2019efficient} for its shorter proofs and faster verification. For a detailed comparison of these schemes, see~\cite{boneh2018survey}, and for the formal definition, refer to Appendix~\ref{app:VDF}.

\noindent \textbf{Aggregate Signatures}: An aggregate signature scheme allows the aggregation of $n$ distinct signatures from $n$ users, each on a distinct message of their choice, into a single signature~\cite{boneh2003aggregate}. Moreover, it allows the aggregation to be done by any party among the $n$ users, including a potentially malicious party. By verifying the aggregate signature, one can be convinced that $n$ distinct users have signed $n$ distinct messages, which have been collected into a single signature. \FIRST{} utilizes this cryptographic primitive to aggregate the verification results of a VDF proof. 

\section{System and Threat Model} \label{sysmodel}

\subsection{System Model}

\textbf{Parties}: In our system, there exist four main entities. 
1) A smart contract \emph{SC} that resides on the Ethereum blockchain. 2) Alice, who is a legitimate user interacting with \emph{SC} by creating a transaction $tx_{A}$ that is potentially vulnerable to frontrunning attacks. Alice is equipped with a verification/signing keypair $(pk_{A}, sk_{A})$. She evaluates a VDF instance, $\VDF$, given to her by a set of verifiers. 3) A set of verifiers $\verifiers$ who generate and send the public parameters of the VDF, $\VDF$, to Alice and verify the evaluated $\VDF$ and its proof of correctness that Alice submits to them. A \emph{coordinator} $\coordinator$, is an entity picked from the members of $\verifiers$ by Alice to help aggregate their signatures into a single signature. 4) Validators, whose goal is to construct blocks and propose them to the network, validate potential blocks received from other nodes, and process transactions. Finally, dApp creator (\dac), who implements  applications such as auctions, exchanges, bug bounty programs, and Initial Coin Offerings (ICOs) which are known to be targeted by frontrunning attacks.


\subsection{Threat Model and Assumptions} \label{adsmodel}
\textbf{Mallory}: We assume Mallory is an adversary who is computationally bounded and economically rational. Mallory is observing the pending transaction pool for Alice's transaction, $tx_{A}$ on the Ethereum network. Mallory will attempt a frontrunning attack as soon as she observes $tx_{A}$ on the pending pool by paying a higher priority fee. We also take into account the case where more than one adversary attempts to frontrun $tx_{A}$. For ease of exposition, we use Mallory to represent a group of adversaries.  \\
\noindent 
\textbf{Verifiers}: Verifiers $\verifiers$ are a set of entities not controlled nor owned by the \emph{dAC}. The protocol in its current version relies on the assumption of an honest majority among verifiers to guarantee the system's proper functionality. The trust assumption in the verifiers ensures that transactions are not subjected to unnecessary delays from the malicious verifiers, thereby maintaining the liveness property. In \FIRST{}, verifiers do not have access to client details during $\VDF$ verification, effectively precluding transaction censorship. 

While we acknowledge the trust assumption in this version, \FIRST{} is designed with an intuitive plug-and-play framework that seamlessly integrates projects like Eigenlayer. This integration aims to minimize trust dependencies and align with Ethereum's renowned fault tolerance~\cite{EIGENLAYER}. Eigenlayer offers Ethereum validators the opportunity to restake their ETH, thereby extending Ethereum's security to additional protocols. Just as with Ethereum's PoS system, any lapse in ensuring protocol security results in a corresponding slash of their stakes. Furthermore, we consider scenarios where a subset of malicious verifiers might attempt to leak transaction details to Mallory, and demonstrate how \FIRST{} prevents such occurrences in Section~\ref{scenarios}.

\noindent \textbf{Validators}:  We assume that the validators are greedy---they sort transactions in descending order of priority fee and pick them in an order that maximizes their profit. It is important to note that in \FIRST{}, verifiers do not have access to client details during $\VDF$ verification, effectively preventing transaction censorship. They can also re-order transactions to increase their profit and attempt to frontrun victim transactions.

\noindent \textbf{Coordinator}: The coordinator is randomly chosen by Alice from a set of verifiers $\verifiers$. It's important to emphasize that while this entity doesn't need to be trusted for security purposes, it is essential for ensuring liveness. We assume Alice actively monitors the transaction process. If any intentional delays are detected, Alice will re-elect a coordinator and continue her interactions with the new entity.  

\noindent \textbf{dApp Creator}: We assume $\dac$ will deploy $SC$ and implement it correctly. We also assume that $\dac$ does not collude with any other participant or with validators as it is in their best interest to protect their dApp for business reasons. Furthermore, the inherent transparency of smart contract code, which is accessible to the public, acts as a safeguard against malicious intent. Moreover, we assume that $\dac$ has both completed the Know-Your-Customer (KYC) process and undergone an audit for the protocol, providing an added layer of deterrence against malicious attempts. While KYC verification is predominantly utilized in centralized services, there are companies like that offer this service for dApps~\cite{solidproof}. KYC verification ensures that in the event of any malicious actions by $\dac$, the real-world entity behind it can be easily identified, thereby enhancing the deterrent effect against potential malicious activities.

We do not discuss networking-related attacks as they are out of the scope of this work; we refer the reader to relevant research~\cite{bdos}. 

\section{The \FIRST{} Framework} \label{construction}
\subsection{Overview of \FIRST{}} \label{descryption}

\revision{The conceptual idea behind \FIRST{} is to prevent Mallory (an attacker) from frontrunning Alice's transaction by ensuring that Mallory’s transaction cannot reach the smart contract before Alice's is posted. To achieve this, \FIRST{} requires every user interacting with a \FIRST{}-protected contract to compute a verifiable delay and wait independently for a predetermined time $t_1$ before submitting their transaction to the mempool. Importantly, \FIRST{} does not impose a global ordering or queue across users; each user's delay is enforced individually. This mechanism guarantees that no transaction becomes visible to adversaries until after the VDF commitment is fulfilled.}

The goal is to choose $t_1$ for a given time period/epoch, s.t. $t_1 >> t_2$, where $t_2$ is the expected wait time of the transaction of any Alice in the mempool before getting posted on the Blockchain. This ensures that, with high probability, Mallory cannot frontrun Alice's transaction that she sees in the mempool. The time $t_2$ depends on several dynamic factors, namely transaction gas price, priority fee, miner extractable value (MEV), and network congestion at the time of submission, which makes an exact assessment of $t_2$ difficult. Since the expected value of $t_2$ is the best can be done, there is a chance of $t_1$ being less than the actual waiting time for Alice's transactions.  Given that $t_2$ is difficult to predict, and a high $t_1$ is detrimental to transaction throughput due to latency, what we do is empirically arrive at a ``reasonable'' value for $t_1$. \FIRST{} continuously monitors the blockchain data to identify the minimum priority fee value that would result in a high likelihood of all \first{} transactions waiting approximately $t_2$ time in the mempool. The $t_1$ wait time is then fixed for a given epoch (higher than $t_2$), ensuring that a potential attack transaction has very low probability to frontrun valid \first{} transactions. For our application of \first{} in Ethereum, we set this epoch to be the same as the default Ethereum epoch of 32 blocks. The $\dac$ obtains the value of $t_1$ via statistical analysis of the relation between the priority fee of the transaction and transaction confirmation time by monitoring the Ethereum network continuously. Consequently, \FIRST{} recommends an optimal priority fee that significantly decreases the likelihood of transactions getting frontrun. 
We detail how we perform such a statistical analysis in Section~\ref{implementation}. 

\vspace{-1.2em}
\subsection{Construction of \FIRST{}}
\label{sec:cons}
\begin{figure}[]
\centering
\includegraphics[width=0.5\textwidth]{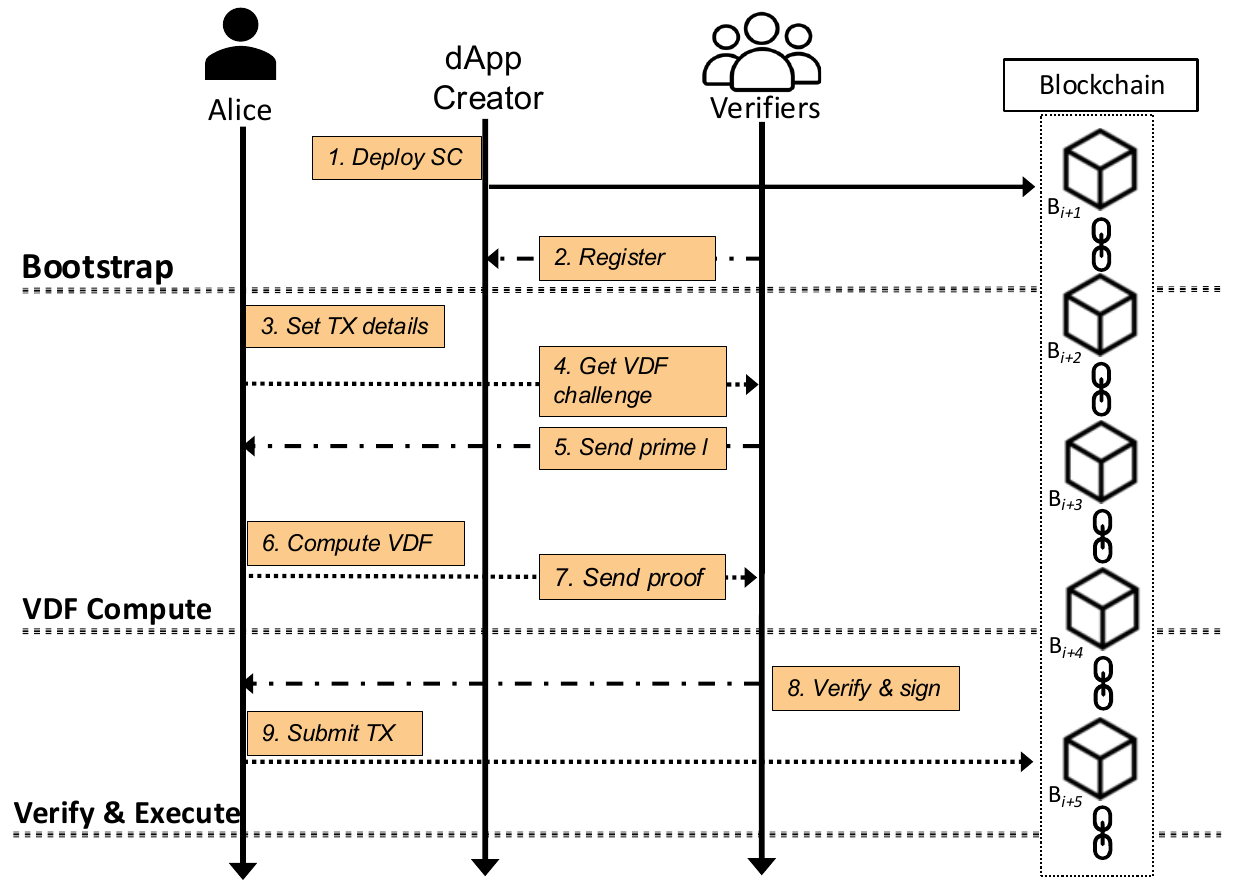}
\caption{Overview of \first{}.}
\label{tbl:exp3_new}
\Description{}
\end{figure}

\revision{This section outlines the key components of \FIRST{}, illustrated in Figure~\ref{tbl:exp3_new}, which consists of seven protocols. 
\textbf{Steps 1–2} correspond to the deployment of the smart contract on the blockchain and the registration of verifiers with the dApp owner. After registration, each verifier independently generates a key pair. These steps constitute the bootstrap phase of the system (Protocols~\ref{proto:SystemSetup} and~\ref{proto:ParamGen2}).
\textbf{Steps 3–4} represent the initialization of a transaction by Alice. She prepares the transaction details and requests a VDF challenge from the verifier set (Protocol~\ref{proto:setup}).
\textbf{Step 5} shows the verifiers responding with a unique VDF challenge—a prime $l$—which Alice must use to compute the VDF (Protocol~\ref{proto:exchange}).
\textbf{Steps 6–8} cover the VDF evaluation and verification phase (Protocol~\ref{proto:proofVerification}). Alice computes the VDF output offchain and sends the proof to the verifiers, who verify its correctness and sign the result.
\textbf{Step 9} corresponds to Alice submitting her transaction, along with the signed proof, to the smart contract on Blockchain (Protocol~\ref{proto:smart:verify}). The contract then verifies both the signature and VDF proof before executing the transaction onchain.
For simplicity, the computation of the recommended transaction tip, $\firsttip$, described in Protocol~\ref{proto:firstCalc}, is not visualized in the figure, but it is applied before the final transaction submission.}



\begin{algorithm}[b]
\SetAlgorithmName{Protocol}{}{}
	\caption{System setup.}
\label{proto:SystemSetup}
	\DontPrintSemicolon
	\SetAlgoLined
	\SetKwInOut{Input}{Inputs}
	\SetKwInOut{Output}{Output}
	\SetKwInOut{Parties}{Parties}
	\Input{Security parameter $\lambda$.}
    \Output{\emph{SC}, $(pk, sk)$ keypair for each member of $\verifiers$.} 
    
    \Parties{dApp creator (\dac), set of verifiers ($\verifiers$).}
        \dac\ implements \emph{SC} and deploys it on Ethereum. 

        Each $V_{i};\, i \in [1\dots n]$ generates $(pk_i, sk_i) \leftarrow \Agg.\KeyGen(1^{\lambda})$.\; 
\end{algorithm} 

We use $\sign$ and $\verify$ with no pre-pended string to denote regular digital signature functions, whereas $\Agg.\text{function}()$ denotes functions specific to the aggregate signature scheme. We use $\verify$ for both, signature and VDF verification, which will be clear from context. Below we discuss each protocol.


\noindent \textbf{Protocol~\ref{proto:SystemSetup}}: This is the bootstrap protocol of \FIRST{}, and is executed only once. It takes a security parameter as input and outputs the smart contract $SC$ and verification/signing keypairs for each member of $\verifiers$. First, the \dac\, implements and deploys the dApp on Ethereum. Entities sign up with the \dac\, to become verifiers.
Following deployment, each member of $\verifiers$ generates their key pairs. \revision{We note that $\leftarrow$ denotes an assignment operation.}

\noindent \textbf{Protocol~\ref{proto:ParamGen2}}: \change{The protocol is used to generate the system parameters of \FIRST{}. It takes in a security parameter and outputs the public parameters ($pp$) of VDF $\VDF$. In Line 1, each  $V_{i} \in \verifiers$ initializes its list $D_{i}$ and $U_{i}$, used to keep track of values used in the VDF $\VDF$'s evaluation and verification, respectively. In Line 2, the dApp creator $\dac$ initializes the number of steps $\Tau$ that will be used in the evaluation of $\VDF$. $\Tau$ is the number of steps required to evaluate the VDF instance which results in a  corresponding delay of $t_1$ units of time. Next, $\dac$ samples a negative prime integer $d$, which satisfies $d \equiv 1 \pmod{4}$. These requirements ensure that when generating the class group~(\text{Cl}) from $d$ in Line 4, the resulting class group order cannot be efficiently computed by any known algorithm~\cite{wesolowski2019efficient,cohen2019chia}. }

Currently, two approaches are known for setting up $\VDF$: using an RSA group of unknown order and using class groups of imaginary quadratic fields~\cite{wesolowski2019efficient} whose order is hard to determine. The RSA group approach requires a trusted setup when generating $N$ such that $N = p.q$ where $p$ and $q$ are primes; in particular, $p$ and $q$ need to be kept secret. On the other hand, a class group of imaginary quadratic fields does not require a trusted setup and is used by blockchains such as the Chia network in production~\cite{cohen2019chia}. We use such groups to eliminate the trusted setup requirement in our construction. 


\LinesNumbered
\begin{algorithm}[t]
\SetAlgorithmName{Protocol}{}{}
	\caption{Parameter generation.}
 \label{proto:ParamGen2}
	\DontPrintSemicolon
	\SetAlgoLined
	\SetKwInOut{Input}{Inputs}
	\SetKwInOut{Output}{Output}
	\SetKwInOut{Parties}{Parties}
	\Input{Security parameter $\lambda$.}
    \Output{$pp$.}
    \Parties{dApp creator (\dac), set of verifiers ($\verifiers$).}
        \change{Each $V_{i} \in \verifiers$, 
        initializes lists $D_{i}, U_{i} = [~]$.}\\
         
         \change{\dac\, picks $T \in \mathbb{Z}^{+}$.}\; 

         \change{ $\dac \sample d$, s.t., $d$ is negative prime and $d \equiv 1 \;(\bmod\; 4)$. }
        
        \change{$\dac$ computes $\mathbb{G} \leftarrow \text{Cl}(d)$ and output  $ pp = (\mathbb{G}, T$)} \; 
        
\end{algorithm} 


\begin{algorithm}[]
\SetAlgorithmName{Protocol}{}{}
	\caption{\first~recommended priority fee calc. 
 \label{proto:firstCalc}
 }
        
	\DontPrintSemicolon
	\SetAlgoLined
	\SetKwInOut{Input}{Inputs}
	\SetKwInOut{Output}{Output}
	\SetKwInOut{Parties}{Parties}
	\Input{Initial $t_2$, $\alpha$, and $k$ (multiplication factor for $t_1$).} 
    \Output{Recommended priority fee.}
    \Parties{dApp creator (\dac).}
\SetKwFunction{FRetFIRSTFee}{$\mathsf{RetrieveFIRSTFee}$}
\SetKwFunction{main}{$\mathsf{main}$}
\SetKwFunction{FCalcFee}{$\mathsf{CalcFIRSTFee}$}
\SetKwProg{Fn}{function}{:}{}
\Fn{\FCalcFee{$\firsttip$,$tx_{list}$}}{
$temp_{list} = [~]$.\\
\For{$tx$ in $tx_{list}$}{
\If{$tx_{wait\_time}  < t_2$}{
$temp_{list}.append(tx_{priority\_fee})$.\\
}
}
\If{$temp_{list}$ == $\emptyset$}{
\tcc{re-calibrate $t_2$ \& $t_1$.}
Initiate epoch change.\\
\Return\\

}
$f_{avg}$ = average$(temp_{list})$.
\\
\If{$\firsttip$ == 0}{
$\firsttip = f_{avg}$.\\
}\Else{
    $\firsttip = \alpha \times f_{avg} + (1 - \alpha) \times \firsttip.$
}

}

\Fn{\main{}}{
\ShowLn
$\firsttip = 0$.\\
\ShowLn
$t_1 = k \times t_2$. \\ 
\ShowLn
\While{True}{
\ShowLn \If{New block with $tx_{list}$ transactions is posted on BC}{
\ShowLn $\mathsf{CalcFIRSTFee}(\firsttip,tx_{list})$.\\
\ShowLn
\If{Current epoch ended}{
\ShowLn
Update $t_2$ if needed, set $t_1 = k \times t_2$ and update $T$ to correspond $t_1$. \\
} } }  }
\end{algorithm}

\LinesNumbered
\begin{algorithm}[t]
\SetAlgorithmName{Protocol}{}{}
	\caption{Transaction detail generation.}
        \label{proto:setup}
	\DontPrintSemicolon
	\SetAlgoLined
	\SetKwInOut{Input}{Inputs}
	\SetKwInOut{Output}{Output}
	\SetKwInOut{Parties}{Parties}
	\Input{$addr_{A}, f_{\text{name}}, addr_{SC}$.}
    \Output{Secret message  $\messages_A$, $h$, Signature $\sigma_{A}$.}
    \Parties{set of verifiers ($\mathbb{V}$), user in system (Alice).}
        Alice generates message $\messages_A = (addr_{A} , f_{\text{name}} , addr_{SC},input_{SC}).$ \\
        
        
        Alice generates hash of $\messages_A$, $h = H(\messages_A)$, and signs it: $\sigma_{A} \leftarrow \sign (sk_{A},h).$ \\
        
        Alice sends $(h,\sigma_{A})$ to each $V_{i};\, i \in [1\dots n], n = |\verifiers|$, including the $V_i$ she picks as the coordinator $\coordinator$ for signature aggregation. 
\end{algorithm}

\noindent \textbf{Protocol~\ref{proto:firstCalc}}: 
The goal of \FIRST{} is to provide users with frontrunning resistant transactions. \revision{To achieve this, we introduce a novel mechanism for computing a custom recommended fee, denoted as $\firsttip$. Without $\firsttip$, users would have to rely blindly on wallet or Etherscan-suggested priority fees or manually estimate gas prices, exposing them to potential delay in transaction inclusion, which increases their vulnerability to frontrunning or may lead to unnecessary overpayment. The $\firsttip$ helps incentivize faster pickup of the FIRST transactions, thus ensuring their faster confirmation and lower latency.
} This protocol continuously monitors blockchain transactions to analyze transaction wait times and the associated priority fees paid, in order to calculate the $\firsttip$. The $\alpha$ value passed as input to the protocol corresponds to the weight of simple Exponentially Weighted Moving Average (EWMA) calculation for $\firsttip$ inside the $\mathsf{CalcFIRSTFee}$ function. 

The value of $t_1$ in Protocol~\ref{proto:firstCalc} corresponds to the $T$ value set in Line~2 of Protocol~\ref{proto:ParamGen2}. $T$ is the estimated number of steps required by a powerful machine (e.g. one with a modern desktop CPU) to compute a VDF proof with $t_1$ delay. Most other less capable machines will take longer than $t_1$ to compute the VDF. For each new block posted on the blockchain, Protocol~\ref{proto:firstCalc} calculates the average fee ($f_{avg}$) paid by transactions which waited less than $t_2$ time in the mempool before being posted in the blockchain. The average value calculated in the previous step is then incorporated into the $\firsttip$ value using EWMA. The \first{} protocol enables the forceful change of an epoch if the $t_2$ value needs to be updated before the current epoch ends. For instance, when the number of transactions in the current block waiting less than $t_2$ time are statistically insignificant or cross a predefined system threshold (in Protocol~\ref{proto:firstCalc}, $temp_{list} == \emptyset$) we initiate epoch change, and update the $t_2$ and $t_1$ values based on the average waiting time across a set number of recent past blocks, which can be a system parameter (10 blocks in our experiments). 

We note that during the shift from a longer to a shorter delay period (t1), \first{} momentarily halts transaction submissions. \revision{This precaution maintains fairness between transactions with varying VDF delays during the transition ($t_1$ to $t^{\prime}_1$), safeguarding transactions with extended VDF delays from being outpaced by those with shorter ones. The pause ensures that all users who started their transaction setup under the old $t_1$ finish their VDF evaluation correctly before the new $t'_1$ becomes active. Without a pause, adversarial users could strategically delay their transaction request to fall after the adoption of $t^{\prime}_1$ to benefit from the reduced delay parameter, which violates fairness guarantees.}

\noindent \textbf{Protocol~\ref{proto:setup}}: This protocol is used to generate transaction details of \FIRST{}'s users. It takes as input Alice's transaction details and outputs a message, its digest and a signature over the digest; the latter two are meant to be given to $\verifiers$. In Line 1, user Alice constructs a tuple, $\messages_{A}$, with the transaction details, including her Ethereum address $addr_{A}$, the dApp smart contract address that she intends to submit a transaction to, $addr_{SC}$ (that \dac\, created), and the name of the function that she intends to invoke to trigger the smart contract $SC$, $f_{\text{name}}$.
We assume she has a verification/signing keypair $(pk_{A}, sk_{A})$, using which, in Line 2, she creates and signs a digest of $\messages_{A}$. Using the cryptographic hash of the transaction details prevents the leakage of any detail that may help a potential frontrunner. Alice sends the digest of $\messages_{A}$ ($h$) and her signature over it ($\sigma_{A}$) to each $V_{i};\, i \in [1 \dots n]$. Alice chooses coordinator ($\coordinator$) from $\verifiers$ to help with signature aggregation in Protocol~\ref{proto:proofVerification} and Protocols~\ref{proto:exchange}.

\begin{algorithm}[t]
\SetAlgorithmName{Protocol}{}{}
	\caption{User-Verifiers interaction.}
 \label{proto:proofVerification}
	\DontPrintSemicolon 
	\SetAlgoLined
	\SetKwInOut{Input}{Inputs}
	\SetKwInOut{Output}{Output}
	\SetKwInOut{Parties}{Parties}
	\Input{$\sigma_{A},h, pp$.}
    \Output{VDF output \emph{y}, VDF proof $\pi$.}
    \Parties{user in system (Alice), set of verifiers ($\verifiers$) including coordinator $\coordinator$.}
        
        
        On receiving $(\sigma_{A}, h)$ from Alice, $\coordinator \in \verifiers$ picks prime $l \sample \mathbb{P}$ and sends $(h,l)$ to $\verifiers \setminus \coordinator$.\; 
        
         

    \For{each $V_{i} \in \verifiers$}
    {
        \If{$l \notin D_i$ and  $l \notin U_i$}
        {
            \If{$\true$ $\leftarrow$  $\verify$($pk_{A},$ $h,$ $\sigma_{A})$}{
                $\messages_i = (l,h, V_{i},block_{curr})$. \\ 
                 $\sigma_{V_{i}} \leftarrow \Agg.\sign(sk_{V_i},\messages_i)$.\;
                 Add $(l)$ to $D_i$.\\ 
                 Send $(M_{i}, \sigma_{V_i})$ to $\coordinator$.

  }
 
}

}    \tcc {Sig. aggregation and evaluate $\VDF$} 
       
        $\coordinator$ checks if each $\Agg.\verify(pk_{i},\messages_{i},\sigma_{V_i}) \overset{?}{=} \true$. If majority of members of $\mathbb{V}$ return $\bot$, $\coordinator$ returns $\bot$ to Alice. Else $\coordinator$ does  $\sigagg$ $\leftarrow$ $\Agg.\aggsign(M_{1}, \dots, M_{j},\sigma_{V_{1}}, \dots,\sigma_{V_{j}})$, where $j > |\mathbb{V}|/2$.

        $\coordinator$ creates $\messages_{\agg} = (\sigagg, M_1, \dots, M_j, pk_1\dots pk_j)$ and sends $\messages_{\agg}$ to Alice. 
        
        Alice checks if $\Agg.\aggverify(\sigagg, M_1, \dots, M_j, pk_1, \dots, pk_j)$ $\overset{?}{=}$ $\true$, and if $\verify (pk_{i}, pp, \sigma_{pp_{i}}) \overset{?}{=} \text{true}$  where $ i \in \{1 \dots j\}$, and $j > |\mathbb{V}|/2$. If both return yes, 
        Alice computes $(\pi, y) \leftarrow \VDF.\eval(pp, l)$. Else returns $\bot$ and retry. 
        
        Alice sends $( \pi, \emph{y})$ to all members of $\verifiers$. 
        
\end{algorithm}

\begin{algorithm}[t]
\SetAlgorithmName{Protocol}{}{}

\caption{VDF verification and tx submission.}
\label{proto:exchange}
\DontPrintSemicolon
\SetAlgoLined
\SetKwInOut{Input}{Inputs}
\SetKwInOut{Output}{Output}
\SetKwInOut{Parties}{Parties}
\Input{$\pi, y$.}
\Output{Aggregate signature $\sigagg^\prime$, transaction $tx_{A}$.}
\Parties{user in system (Alice), set of verifiers ($\verifiers$).}

\tcc{Each verifier runs proof verification}
\For{each $V_{i} \in \verifiers$}
{
    \If{$l$ $\notin$  $U_i$ \text{ and } $l \in D_i$}
    {
        Add $l$ to $U_i$.\\
        \If{``accept'' $\leftarrow$  $\VDF.\verify$(pp, l, \emph{y}, $\pi$)}
        {   
            $M^{\prime}_{i} = (``accept", V_{i}, l) $. \\
        
            $\sigma_{V_{i}}^{\prime} \leftarrow \Agg.\sign(sk_{V_i}, M^{\prime}_{i})$. \\
            
            Send $(M^{\prime}_{i}, \sigma^\prime_{V_{i}})$ to $\coordinator$.

        }
       
    }
    
}  

\tcc{Sig. aggregation and submit tx.}


$\coordinator$ checks if each $\Agg.\verify( pk_{i}, M^{\prime}_{i}, \sigma^\prime_{V_i})  \overset{?}{=} \true $. 
If majority of members of $\mathbb{V}$ return $\bot$, $\coordinator$ returns $\bot$ to Alice. Else $\coordinator$ does $\sigagg^{\prime}$ $\leftarrow$ $\Agg.\aggsign(M_{1}^{\prime}, \dots, M_{j}^{\prime}, \sigma_{V_{1}}^{\prime},\dots,\sigma_{V_{j}}^{\prime})$, where $j>|\mathbb{V}|/2$.

$\coordinator$ creates $M_{\agg}^{\prime} = (\sigagg^{\prime}, M_{1}^{\prime},\dots,M_{j}^{\prime}, pk_{1},\dots,pk_{j})$ and sends to Alice.\\

Alice checks if $\Agg.\aggverify(\sigagg^{\prime}$, $M_{1}^{\prime},$ $\dots, M_{j}^{\prime},$ $pk_1, \dots, pk_j)$ $\overset{?}{=} \true$ and $j > |\verifiers|/2$, if yes, Alice creates $M^\prime = (\messages_A, \messages_{\agg}, M_{\agg}^{\prime})$ and signs it, $\sigma_{A}^{\prime} \leftarrow \sign(sk_{A}, M^{\prime})$. Else returns $\bot$ and retry. \\ 

Alice retrieves the current recommended priority fee ($\firsttip$) from Protocol~\ref{proto:firstCalc}.

Alice creates and submits transaction $tx_A = (\sigma_{A}^{\prime}, M^{\prime}, pk_{A}, \firsttip).$ 

\end{algorithm}

\begin{algorithm}[t]
\SetAlgorithmName{Protocol}{}{}
	\caption{Signature validation and SC execution.}
        \label{proto:smart:verify}
	\DontPrintSemicolon
	\SetAlgoLined
	\SetKwInOut{Input}{Inputs}
	\SetKwInOut{Output}{Output}
	\SetKwInOut{Parties}{Parties}
	\Input{$tx_{A}$, $block_{now}$ and $threshold$.} 
    \Output{Smart Contract Functionality.}
    \Parties{User in system (Alice), Smart Contract (\emph{SC}).}
    
    Parse $tx_A = (\sigma_{A}^{\prime}, M^{\prime}, pk_{A}, \firsttip)$, $M^\prime = (\messages_A, M_{\agg}, M_{\agg}^\prime)$, $\messages_{\agg} = (\sigagg, M_{1},\dots,M_{j}, pk_{1},\dots,pk_{j})$, and $M_{\agg}^\prime = (\sigagg^{\prime}, M_{1}^\prime,\dots,M_{j}^\prime, pk_{1},\dots,pk_{j})$, where $j > |\mathbb{V}|/2$.

    \If{ ($H(addr_A, f_{name}, addr_{SC},input_{SC}) \overset{?}{=} h$) and ( $(l,h,\cdot, block_{curr}) \in [M_{1},\dots,M_{j}]$) and (($``accept",\cdot,l) \in [M_{1}^\prime,\dots,M_{j}^\prime]$) and ($|M_{1},\dots,M_{j}| > |\mathbb{V}|/2$) and ($|M_{1}^\prime,\dots,M_{j}^\prime| > |\mathbb{V}|/2$)} {
        
         \If{$\Agg.\aggverify(\sigagg^{\prime}$, $M_{1}^{\prime},$ $\dots, M_{j}^{\prime},$ $pk_1, \dots, pk_j)$ $\overset{?}{=} \true$}
            {
            
                \If{$block_{now} - block_{curr} < threshold$} {
                         \emph{SC} executes the intended functionality.
                    }   
            }
    }
\end{algorithm}

\noindent \textbf{Protocol \ref{proto:proofVerification}}: This protocol must be executed between Alice and members of $\verifiers$. It takes as input the output of Protocol~\ref{proto:setup}, 
i.e., the digest/signature over Alice's message. It  outputs the evaluation of the VDF instance, $\VDF$, and its corresponding proof.  
In Line~1, the coordinator $\coordinator$ samples 
a unique (per user) prime $l$ from a set of primes $\mathbb{P}$ that contains the first $2^{2\lambda}$ primes. 
We require each $V_i$ to independently check $l$ and verify that it was not generated before (Lines 2, 3). 

Upon checking the validity of $l$ and Alice's signature, each $V_i$ creates a message $\messages_{i}$ by concatenating $l$, $block_{curr}$, and $h$ from Protocol~\ref{proto:setup} to its id and signs $\messages_i$ (Lines 5, 6). $\VDF$'s freshness $block_{curr}$, which represents the block height at the time of request is included in $\messages_{i}$ to prevent off-line attacks on $\VDF$. For an off-line attack, Mallory requests $l$ and pre-evaluates the $\VDF$ to submit the frontrunning transaction when the victim transaction is seen on the network. However, the smart contract eliminates this attack by verifying the freshness of $\VDF$.
In Line 7 and 8, each $V_{i} \in \verifiers$ updates their $D_{i}$ to keep track of used $l$ values and sends their $\sigma_i$ to $\coordinator$. The list $D_i$ is used to ensure that no user in the system has been given the current $l$ for $\VDF$ computation, else colluding users can reuse proofs. The list $U_i$ is used to ensure that users in the system can only use a given $l$ once, hence thwarting any replay attacks. 
In Lines 18 and 19, $\coordinator$ verifies the signatures of verifiers, aggregates them, and sends the aggregate signature to Alice for verification. We note that both $D$ and $U$ are public lists.

The goal of the aggregate signature scheme in \FIRST{} is to cut down the cost of verifying each $V_i$'s signature individually. 
Moreover, we can obtain aggregate signatures from all members of $\verifiers$ without requiring any trust assumption on them. We refer interested readers to~\cite{boneh2003aggregate} for further details on the aggregate signature scheme. Alice checks the validity of $\sigagg$ and the number of received messages $j$, where $j > |\mathbb{V}|/2$. If both return true, Alice retrieves and verifies the public parameters of $\VDF$, $pp$, and starts the evaluation of $\VDF$ (we recollect that per our system model, Alice evaluates $\VDF$). During the evaluation, Alice generates output and proof of correctness $\pi$, which is sent to all members of $\verifiers$ (Lines 20, 21).

\noindent\textbf{Protocol~\ref{proto:exchange}}: Protocol \ref{proto:exchange} is required to be executed between Alice and $\verifiers$. It takes as input the VDF evaluation result and its proof (given as output by Protocol~\ref{proto:exchange}), the $pp$ of $\VDF$ and outputs Alice's transaction $tx_A$ to be submitted to $SC$. 
In Line 2, every $V_{i} \in \verifiers$ first checks if $l \notin U_i$. This check ensures that Mallory is not reusing the $l$ to evaluate $\VDF$. Each $V_i$ also checks if $l \in D_i$, to check if $l$ has indeed been assigned to a user. If the check returns true, every $V_{i} \in \verifiers$ adds $l$ to $U_i$. In Line 4, every $V_{i} \in \verifiers$ verifies the VDF proof $\pi$ sent by Alice. Depending on the outcome of the verification, each $V_{i}$ creates $M_{i}^{\prime}$ and signs it (Lines 5, 6). 
Upon completion of the verification phase, in Line 17, $\coordinator$ first verifies each $\sigma_{i}^\prime$ and aggregates the signatures into a unique signature $\sigagg^\prime$. In Line 18, $\coordinator$ creates a tuple $M_{\agg}^{\prime}$, containing the $\sigagg^{\prime}$, distinct messages of members of $\verifiers$, their public keys, and sends it to Alice. Alice checks the validity of $\sigagg^\prime$ and the number of received messages  $j$, where $j > |\mathbb{V}|/2$,  for a majority of verifiers from $\verifiers$. \footnote{The number of messages received by Alice in Protocol \ref{proto:exchange} and Protocol \ref{proto:proofVerification} are both denoted by $j$, but we note that the value of $j$ in both protocols need not be exactly the same, as long as it satisfies the property  $j > |\mathbb{V}|/2$.}
If both return true, Alice creates $M^\prime$ consisting of her message, $\messages_A$ from Protocol~\ref{proto:setup}, $M_{\agg}$ from Protocol~\ref{proto:exchange}, and $M_{\agg}^\prime$ from Protocol~\ref{proto:proofVerification}. Alice signs it before creating transaction $tx_A$, sets the transaction fees ($\firsttip$ from Protocol~\ref{proto:firstCalc} and current Ethereum base fee), and submits the transaction (Lines 19, 20, 21). 



\noindent\textbf{Protocol~\ref{proto:smart:verify}}:This protocol is used to validate the transaction $tx_A$, $\VDF$'s verification details, and the signature aggregation. Alice creates and submits $tx_{A}$ with recommended fee. \emph{SC} parses $tx_{A}$ to access necessary fields. $SC$ verifies transaction details committed to in Protocol~\ref{proto:setup}, verifies the messages of verifiers and checks if the number of participants in the verification phase is more than $|\mathbb{V}|/2$. Finally, \emph{SC} will check the given $\VDF$'s freshness by checking if the difference between the block height at the time of request and the current lies within a pre-defined system threshold that should be adjusted by $\dac$. We note that SC examines all messages and employs the $block_{curr}$ value endorsed by the majority to validate the freshness of VDF, rather than relying on single $block_{curr}$. The \emph{SC} will abort the function execution if any check fails. 


\section{Security Analysis of \FIRST{}}\label{scenarios}
\subsection{Informal Security Analysis}
In this section, we analyze the security of \FIRST{} informally by considering potential attack scenarios and describe how \FIRST{} eliminates them.\\
\noindent \textbf{Malicious Verifier:} In this attack, an adversary might try to corrupt some members of $\verifiers$, and try to glean information about Alice's transaction $tx_A$ while she computes the VDF. \FIRST{} accounts for this by having Alice conceal all transaction details by hashing them and sharing only the digest with the verifiers (Protocol~\ref{proto:setup}, Steps 2-3), thus preventing any leakage of sensitive information. The general security guarantees apply for the case where a malicious verifier attempts to frontrun Alice.

\noindent \textbf{Proactive Attacker:} Consider a scenario where Mallory or a bot she created is monitoring the pending transaction pool to identify a  transaction $tx_{A}$ submitted by a user Alice.
Let $t_{M}$ represent the time Mallory first sees $tx_{A}$, with a gas price $G_{A}$, on the pending pool. Mallory creates a transaction $tx_{M}$ with gas price $G_{M}$ where $G_{M} > G_{A}$. We note that, in order for this attack to succeed, $tx_{M}$ is required to be included in the previous or in the same block but before $tx_{A}$. To address this, \FIRST{} assigns $\VDF$ related parameters and updates them regularly using the empirical analysis we describe in Section~\ref{implementation}. Since all valid transactions need to wait for \first{} stipulated time delay ($\VDF$ delay), $tx_M$ will need to wait to generate valid $\VDF$ proof. If Alice paid the \first{} recommended priority fee, $tx_A$ will wait for at most $t_2$ time in the pending pool, and since $t_2$ is less than the $\VDF$ delay set by \first{} ($t_1$), Alice's transaction will not get frontrun by Mallory with high probability.

\noindent \textbf{Backrunning and Sandwich attack:} 
Backrunning is another attack strategy where Mallory creates a transaction $tx_{M}$ with a gas price of $G_{M}$ where $G_{M} < G_{A}$ to take advantage of the outcome of Alice's transaction~\cite{qin2021quantifying}. Given the enforced $\VDF$ delay, the malicious transaction attempting to backrun the victim transaction has 
to wait before entering the mempool, which prevents backrunning, making it impossible for the attacker's transaction to be scheduled in the same block thus preventing frontrunning. Given both frontrunning and backrunning are prevented, sandwich attack is also prevented~\cite{zhou2020high}.
 
 \noindent \textbf{Malicious Block Proposer:} In Ethereum 2.0, the block proposers are randomly chosen from active validators whose aim is to propose a potential block for the slot they are assigned. As a result, they have full control over inserting, excluding, and re-ordering transactions akin to miners in the PoW version of Ethereum. 
A potential attack can be frontrunning transaction inserted into the block by the malicious block proposer. However, \FIRST{} already handles this case: if any transaction does not contain the aggregated signature of $\verifiers$ on the verification of $\VDF{}$ proof, the smart contract will reject the transaction.

\noindent\revision{\textbf{Impact on Blockchain Throughput: } In EVM-based blockchains, throughput is constrained by the block gas limit—30 million gas on Ethereum—with new blocks produced roughly every 12 seconds. Since VDF computation and verifier coordination are performed entirely off-chain, only the final, verifier-signed transaction is submitted on-chain. As a result, FIRST does not affect blockchain throughput. Once submitted, a FIRST transaction behaves like any other following the standard inclusion and confirmation processes.}

\noindent \textbf{Pre-computed VDF attack:} Attackers may attempt to create frontrunning transactions in advance and broadcast them when they see the victim transaction in the pending pool.  
We eliminate this pre-computation attack vector by checking the freshness of the VDF during smart contract execution (Line~4, Protocol~\ref{proto:smart:verify}). Specifically, suppose the difference between the current block (where the transaction is slated for execution) and the block height at the time of the transaction request is greater than a pre-defined system threshold, the transaction will be reverted. \revision{Verifiers may optionally charge a small fee for issuing a new prime $l$ for the VDF to create an economic disincentive against repeated VDF challenge requests. Since verifier interaction occurs offchain, standard Web2 techniques—such as IP-based rate limiting or user-level quotas—can also be applied to mitigate abuse.}

\subsection{Formal Security  Analysis}
 \label{sec:uc}

We analyze the security of \FIRST{} in the Universal Composability (UC) framework~\cite{canetti2004universally}. To this end, we define an ideal functionality, $\F{FIRST}$, consisting of three functionalities, $\F{setup}$, $\F{bc}$, and $\F{construct}$ along with two helper functionalities $\F{sig}$~\cite{canetti2004universally} and $\F{vdf}$~\cite{landerreche2020non}. We assume that the optimal functionalities share an internal state and can access each other's stored data. We prove the following theorem in the Appendix~\ref{UCFUNC}.


 

\begin{theorem}
\label{thm:uc}
Let $\F{\first}$ 
 be an ideal functionality for \FIRST{}. Let $\adv$ be a probabilistic polynomial-time (PPT) adversary for \FIRST{}, and let $\simu$ be an ideal-world PPT simulator for $\F{\FIRST}$. \FIRST{} UC-realizes $\F{\first}$ for any PPT distinguishing environment $\env$.
\end{theorem}

\begin{figure*}[]
\begin{mdframed}
\begin{center}
{\textbf{Functionality} $\F{bc}$}
\end{center}
    \textbf{Miner $p_i$ requesting current $d_i$}: Upon receiving $(RequestRound,sid)$ from $p_i$, send $d_i$ to $p_i$.
    
    \textbf{Adversary corrupting Miner $p_i$}: Upon receiving $(corrupt, p_i,sid)$ from $\adv$, if $ |~ \mathbb{H} \setminus \{p_i\} ~ | > \mathbb{P}/2$ then set $\mathbb{H} := \mathbb{H}\setminus \{p_i\}$, else return $\bot$.  \textbf{Block hashing}: When $\ctrTime == \blockHashTime$, $\F{bc}$ takes a set of tuples $\mathbb{TX}_B$ such that $\mathbb{TX}_B \subseteq  \mathbb{TX}_P$ where $\mathbb{TX}_P$ represents the set of transactions in $\txpooltable$, and $|\mathbb{TX}_B| = l$. $\mathbb{TX}_B$ is picked such that $l = min(|\mathbb{X}|, \forall \mathbb{X} \in \mathcal{P}(\mathbb{TX}_P))$ and $ \sum_{i=1}^{l} tx_i.txfee \lessapprox \bcblockmaxfee$  where $\mathcal{P}$ represents a power set function. $\F{bc}$ then adds $\blocknum$ to each tuple (e.g. tuple $(tx,\blocknum) ~\forall~ tx \in \mathbb{TX}_B$) and moves them to $\bctable$ and sends to $\simu$ and $\adv$. $\F{bc}$ sets $\ctrTime = 0$ and $\blocknum = \blocknum + 1$.

    \textbf{BC data request handling}: 
    $\F{bc}$ on receiving request $(\getData,sid)$ from user $u$, retrieves all data tuples from $\bctable$, $\txpooltable$, and $\sctable$, and sends to $u$, $\simu$.

    \textbf{BC block num request handling}: 
    $\F{bc}$ on receiving request $(\getBlockNum,sid)$ from a user return $\blocknum$.

    \textbf{Initialization of $\bc$:}
    On receiving $(init,sid, \mathbb{P},\mathbb{H},\bcblockmaxfee,\blockHashTime)$ from $\env$, initialize for each BC miner/validator $p_i \in \mathbb{P}$ a bit $d_i
    := 0$, sets $\bcblockmaxfee$ as the max fee limit for each block, sets current block hashing interval time as $\blockHashTime$, sets $\ctrTime=0$,  and set $\blocknum = 0$. Set $\mathbb{H} \subset \mathbb{P}$ to be set of honest validators.

    \textbf{Smart contract deployment}: $\F{bc}$ on receiving $(sid,\deploy,SC.id,code)$ from any node stores the tuple $(SC.id,code)$ in an $\sctable$ 
    for later retrieval and execution. The $code$ of $SC.id$ will eventually call $\F{setup}$ to verify the hash of $M_u$ in $\hashtable$ and the aggregate signature in $\stable$ of some submitted transaction $tx = (M_u,\sigma_{u},SC.id,(\sigagg$, $M_{1},\dots M_{n}$, $pk_{1}, \dots, pk_{n}), txfee)$ and check that majority of $M_{1},\dots,M_{n}$ contain an $(``accept",\cdot,\cdot)$. If verification fails, then $SC$ outputs a failure $tx^{\prime}$, else it continues execution of the $SC.id$ $code$ which will include verifying hash $h$ of the submitted transaction $M_u$, and finally outputs a successful $tx^{\prime}$.

     \textbf{Transaction request handling}: 
    $\F{bc}$ on receiving $(sid,\invoke,tx)$ 
    stores the $tx$ tuple in $\txpooltable$. If the $tx$ is invoking a smart contract $SC.id$, then $\F{bc}$ retrieves the tuple $(SC.id,code)$ from $\sctable$. $\F{bc}$ executes $code$ with the given $tx$ and the output transaction $tx^{\prime}$ is generated. Both transactions are added to $\txpooltable$ and also sent back to user $u$ and $\simu$. All rows in $\txpooltable$ are arranged in descending order of $txfee$ at all times. 

    \textbf{Miners stepping the time counter forward}: Upon receiving message $(RoundOK,sid)$ from party $p_i$ set $d_i:= 1$. If for all $p_j \in \mathbb{H} : d_j = 1$, then reset $d_j := 0$ for all $p_j \in \mathbb{P}$ and set $\ctrTime = \ctrTime + 1$. In any case, send $(switch, p_i)$ to $\adv$. The adversary is notified in each such call to allow attacks at any point in time.

\end{mdframed}

\caption{Ideal functionality for blockchain.}
\label{fig:fbcp1}
\Description{}
\vspace{-0.1in}

\end{figure*}

\begin{figure}[]
\begin{mdframed}
\begin{center}
\textbf{Functionality} $\F{construct}$ 
\end{center}

\textbf{User request}: Upon receipt of tuple (sid, $\mathsf{req}$, $h, \sigma_{u}$) from a user $u$ with identifier $uid$, $\F{construct}$ adds $(uid, h, \sigma_{u})$ to $\utable$, and returns ``success'' to $u$, and forwards $(sid, \mathsf{req}, uid, h, \sigma_{u})$ to $\verifiers$ and $\simu$. 

\textbf{User response}: Upon receiving $(sid, \text{aggregated},\sigagg, uid)$ from $\coordinator$, $\F{construct}$ looks for a tuple $(\sigagg, \cdot, \cdot)$ in $\stable$; if such a tuple exists $\F{construct}$ retrieves tuple $(l,uid,\cdot)$ from $\ctable$, constructs and returns tuple $(sid, l, \sigagg,\cdot,\cdot)$ to $uid$ and to $\simu$, else returns $\bot$ to both. 

\textbf{User verification}: Upon receiving ($sid, \mathsf{verify}, l, p, s$) from \emph{uid}, $\F{verify}$ checks if $(l,uid,\text{``not-used''})$ exists in $\ctable$; if yes, updates tuple in $\ctable$ to $(l,uid, \text{``used''})$ and forwards ($sid, l, p, s$) to each $V_{i} \in \set{V}$ and $\simu$, else $\F{verify}$ returns $\bot$ to Alice and $\simu$.

\textbf{Coordinator request}: Upon receipt of a message $(sid, l, uid)$ from $\coordinator$, $\F{construct}$ checks if there exists a tuple $(uid, h, \sigma_{u})$ in $\utable$. If not, return $\bot$ to $\coordinator$ and $\simu$. If $(l, \cdot, \cdot)$ exists in $\ctable$, return $(sid, \text{fail}, l)$ to $\coordinator$ and $\simu$, if $(l, \cdot, \text{``used''})$ already exists in $\ctable$, return $(sid, \text{used}, l)$ to $\coordinator$ and $\simu$. Else, $\F{construct}$ 
adds $(l,uid, \text{``not-used''})$ to $\ctable$, $\F{construct}$ retrieves $(u, pk_{u})$ from $\idtable$, constructs tuple $(sid, \mathsf{valid}, l, h, \sigma_{u}, pk_{u})$ and forwards to all $V_{i} \in \set{V}$ and $\simu$. 

\textbf{Coordinator response}: Upon receiving $(sid, V_{i}, m_{i}, \sigma_{V_{i}})$ from members of $\verifiers$, $\F{construct}$ forwards $(sid, V_{i}, m_{i}, \sigma_{V_{i}})$ to $\coordinator$ and $\simu$.
\end{mdframed}

\caption{Ideal functionality for transaction processing and VDF construction.}
\label{fig:fcons}
\Description{}
\end{figure}

\section{Experimental Results and Analyses} \label{implementation}
We evaluate the performance of \FIRST\ on real Ethereum traces over a month long period of observation.
We analyze \first's suggested $\firsttip$ during our experiment and show the effectiveness of \first{} in terms of the percentage of frontrunnable transactions in a given time period. 

A low percentage implies that transactions submitted during the said time period with $\firsttip$ are seldom frontrun. The success of \first{} is not only dependent on the \first{} system parameters, namely $k, alpha,$ and $t_2$, but also on the system-specific network dynamics. We replicate our analysis of \FIRST{} over a non-EIP-1559 chain, Binance Smart Chain (BSC). In what follows, we discuss details of our experimental setup, data gathering, and experimental results.

\subsection{Data Gathering}
\label{sec:dataagg}
In order to get the most accurate waiting times of transactions in the pending pool, we deployed a  Geth\footnote{https://github.com/ethereum/go-ethereum} full node (v.1.11.0) running on an Amazon AWS Virtual Machine located in North Virginia. The AWS node had an AMD EPYC 7R32 CPU clocked at 3.30 GHz with 8 dedicated cores, 
32 GB of RAM, 1.3 TB solid-state drive, running Ubuntu (v.$20.4$). We also ran a beacon node using  Prysm\footnote{https://github.com/prysmaticlabs/prysm} (v.3.1.2) software which is required to coordinate the Ethereum proof-of-stake consensus layer operations. Once the deployed node synced, we collected the data in the Geth node's pending pool. The data collected included transaction arrival times and the transactions' corresponding unique transaction hashes from block number 15665200 to block number 15886660\footnote{\url{https://web3js.readthedocs.io/en/v1.2.11/web3-eth-subscribe.html}}. For each collected transaction from the confirmed blocks on the blockchain, we gathered additional details such as block base fee, paid max priority fee, gas price, and block confirmation time. Although the data is from October 2022, the transaction volume has remained stable since then (see \cite{etherscanChart}), and Ethereum still lacks frontrunning protection at the application layer, making the data relevant.


We used a machine with Apple M1 Max chip, $32$~GB RAM, $1$~TB HDD, running macOS Monterey (v.$12.6$) to perform experiments on the collected data. 
There were a total of 30.6M transactions for the given block range (15665200--15886660), out of which, 24.34M were Type-2 (EIP-1559) transactions and 6.26M were Type-1 (legacy, non-EIP-1559) transactions. 
We analyzed the more common Type-2 transactions, \first{} can also be used for Type-1 transactions. Out of the total 30.6M transactions our node was able to detect the wait time for 29.65M transactions. Since our node did not receive a total of 944807 transactions (roughly $3.08\%$), we conclude that these transactions were either never sent to the P2P layer because of the use of relayers (e.g., Flashbots) or our node did not receive them before their confirmation on the blocks due to network latency. In practice, the dApp owner would deploy multiple full nodes to collect the pending pool data, hence minimizing the chance of missing transactions due to network latency. We deployed another full Geth node in AWS in Singapore with the same software and hardware specifications as the one in North Virginia. The intent was to perform a comparative sanity-check on the transactions copies recorded at two geographically diverse locations.

We computed the waiting times of transactions received by our node by subtracting the transaction's block confirmation time from the recorded time when the transaction was first seen in our node's pending pool. 
The difference in waiting times of transactions in the US and Singapore was very small. 
Across all the transactions that we captured, the difference between the receipt times in the US and Singapore was no more than 2 ns for any transactions. Interestingly, our Singapore node also never received any of the 944807 transactions that were not seen by our US node, leading us to conclude that those transactions were privately relayed.


\subsection{Extension to non-EIP-1559 chain}
\label{sec:noneip}

Many Ethereum Virtual Machine (EVM) based blockchains, such as Polygon and Fantom, have implemented the EIP-1559 patch. Despite the overall trend of EVM-based blockchains adopting EIP-1559, for completeness, we also studied a non-EIP-1559 chain protocol. We replicated our analysis on the Binance Smart Chain (BSC) which is currently a non-EIP-1559 chain. We deployed a Geth node (v.1.1.17) on AWS Singapore and recorded transaction wait times for 45K blocks (23285229--23288229), totaling 5.29M transactions (statistically significant). Out of the 5.29M transactions, our node did not receive 141157 ($2.66\%$).  
In non-EIP-1559 chains, the gas price is used to incentivize the validator to pick up a transaction. Hence, \first{} uses gas price to calculate the $\firstfee$ in Protocol~\ref{proto:firstCalc}.

\subsection{Aggregate Signature Implementation}
To assess the cost associated with verifying \FIRST{} transactions via a smart contract, we implemented the aggregated signature verification function~\cite{boneh2003aggregate} using the Solidity programming language and deployed it within a smart contract.

\begin{table}[]
\centering
\caption{GAS CONSUMPTION FOR AGGREGATE SIGNATURE VERIFICATION ON SMART CONTRACT. }
\label{gasTable}
\small
\begin{tabular}{c|c}

\textbf{Number of Verifiers} & \textbf{Total Gas Consumption (gas units)} \\ \hline

5  & $374423$ \\\hline
7 & $474327$ \\\hline
10 & $621180$\\\hline
15  & $871987$\\\hline
20  & $1122943$\\\hline
\end{tabular}
\end{table}

We used elliptic curve pairing operations, such as addition, multiplication, and pairing checks introduced by Ethereum in the form of precompiled contracts with EIP-197~\footnote{https://eips.ethereum.org/EIPS/eip-197}. In Ethereum, precompiled contracts enable the deployment of computationally-intensive operations at a lower cost compared to the users implementing them on their smart contracts. Our implementation uses the \texttt{alt\_bn128} curve. We used the bn256\footnote{https://pkg.go.dev/github.com/cloudflare/bn256} library (v.0) and the Go programming language (v.1.17.5) to implement the aggregate signature generation and verification schemes. Table~\ref{gasTable} shows our results for the verification of the aggregated signatures by the smart contract with different numbers of verifiers. For example, it costs 621180 units of gas for ten verifiers to verify the aggregated signature. 
Using the median gas cost of \textbf{10 GWei} (representative of current market conditions as of December 2024) and the current Ether price of \textbf{\$3300}, the cost to verify the aggregated signature of 7 verifiers is approximately \textbf{\$15.65}.



\revision{A 2021 study found that frontrunning extracted approximately \$18 million across 200{,}000 transactions, or about \$90 per transaction~\cite{torres2021frontrunner}. \FIRST\ adds an overhead of \$10--\$15 per transaction, mainly due to the on-chain signature verification. This cost represents a 15\% premium per protected transaction. 
The primary contributor to the cost is the pairing operations required for signature verification. Currently, \textbf{EIP-197} is the only supported precompiled contract for pairing operations on Ethereum, using the \texttt{alt\_bn128} curve. This limited support contributes to the high gas costs. While proposals like \textbf{EIP-2537} aim to introduce more efficient cryptographic primitives such as \texttt{BLS12-381}, they have not yet been implemented on Ethereum mainnet as of May 2025. Once adopted, these upgrades are expected to significantly reduce verification costs. In addition, designing efficient aggregate-signature schemes is an area of continuing research---we anticipate that the schemes to become more efficient in the future.}

\revision{\FIRST\ serves as a proof-of-concept demonstrating frontrunning protection on EVM-compatible chains. Additional savings are possible using \emph{gas golfing}\footnote{Gas golfing refers to low-level optimizations such as inline assembly.}. We will assess that in the future. A note of caution is that inline assembly is error-prone and a known source of vulnerabilities in smart contracts~\cite{chaliasos2022study}.}

\subsection{Scalability of VDF}
\label{sec:scalability}
 
We assess the practicality of VDF on devices with varying computational capabilities by comparing the VDF computation times on these devices. The VDF~\cite{wesolowski2019efficient} used in FIRST has a complexity of $O(T)$ for VDF proof generation and $O(\log T)$ for verification, where $T$ is the number of steps required for proof generation. 
\begin{figure} [b]
    \includegraphics[width=0.45\textwidth]{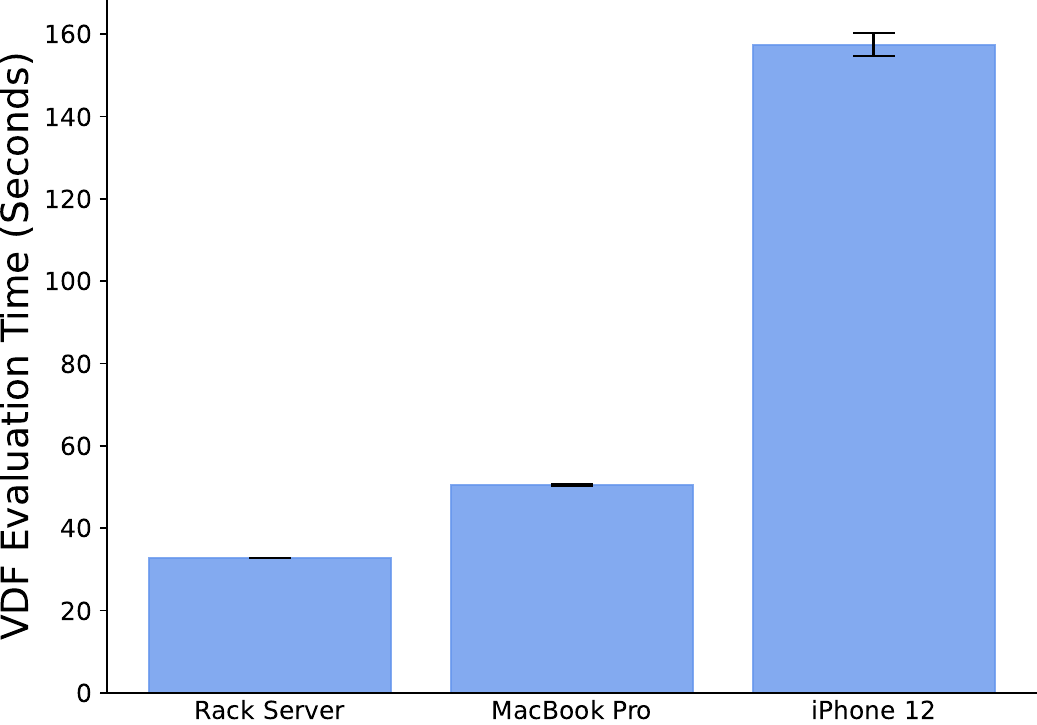}
     \caption{Comparison of VDF computation times across multiple devices.}
        \label{fig:vdfDiff}
     \vspace{-0.1in}
     \centering
     \Description{}
\end{figure}

For client-side costs experiments, we use a rack server, specifically the PowerEdge R650 Intel Xeon Gold 6354 with 18 cores and 36 threads per core, equipped with 256 GB RDIMM and NVIDIA Ampere A2. Additionally, we evaluated the performance on an iPhone 12 with an A14 Bionic 6-core CPU, 64GB storage, and 4GB RAM, as well as the MacBook Pro. Detailed specifications for the MacBook Pro used in our evaluation can be found in Section~\ref{sec:dataagg}. For this experiment, we choose the sequential steps amount (T) to be 1 million whereas the bit length of security parameters to be 2048 bits. We give our results in Figure~\ref{fig:vdfDiff}. \revision{While Figure~\ref{fig:vdfDiff} shows that rack servers compute VDFs faster than other devices, the difference is not sufficient to enable successful frontrunning. This is because no entity, regardless of computational resources, can see the transaction details until after the user's VDF computation is complete and the transaction enters the pending pool. Even highly resourceful devices do not have enough time to complete a new VDF computation fast enough to frontrun, especially when the user includes the optimized priority fee recommended by $\firstfee$.}

\subsection{Analyses and Discussion}
\label{sec:analysis}

\begin{figure*}[!htbp] \label{tbl:FirstFEE}
  \centering
  \subfloat[Recommended FIRST Fee for Ethereum.]{
  \includegraphics[width=0.45\textwidth]{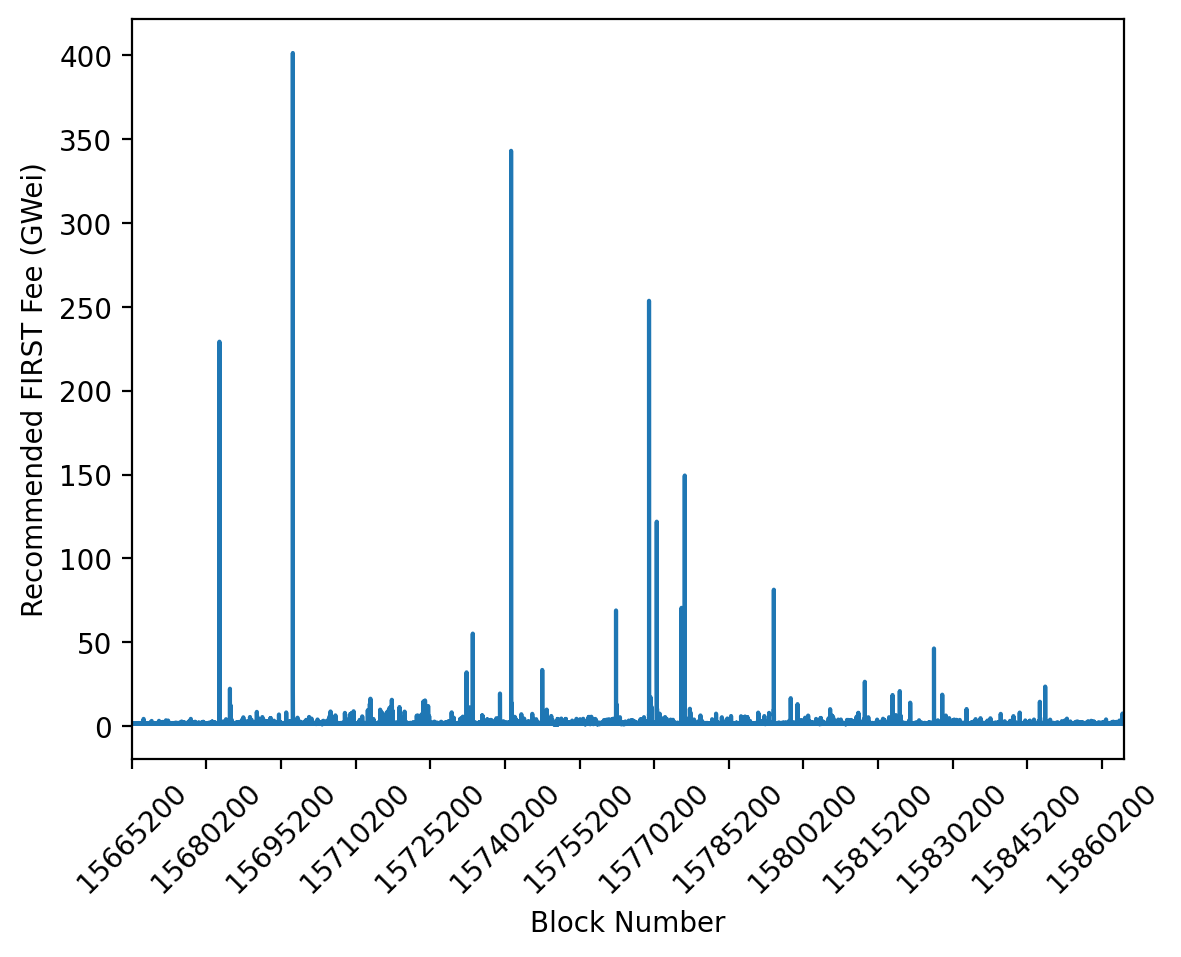}\label{FirstFEEa}}\quad 
  \subfloat[Recommended FIRST Fee for BSC.]{
    \label{FirstFEEb}
  \includegraphics[width=.43\textwidth]{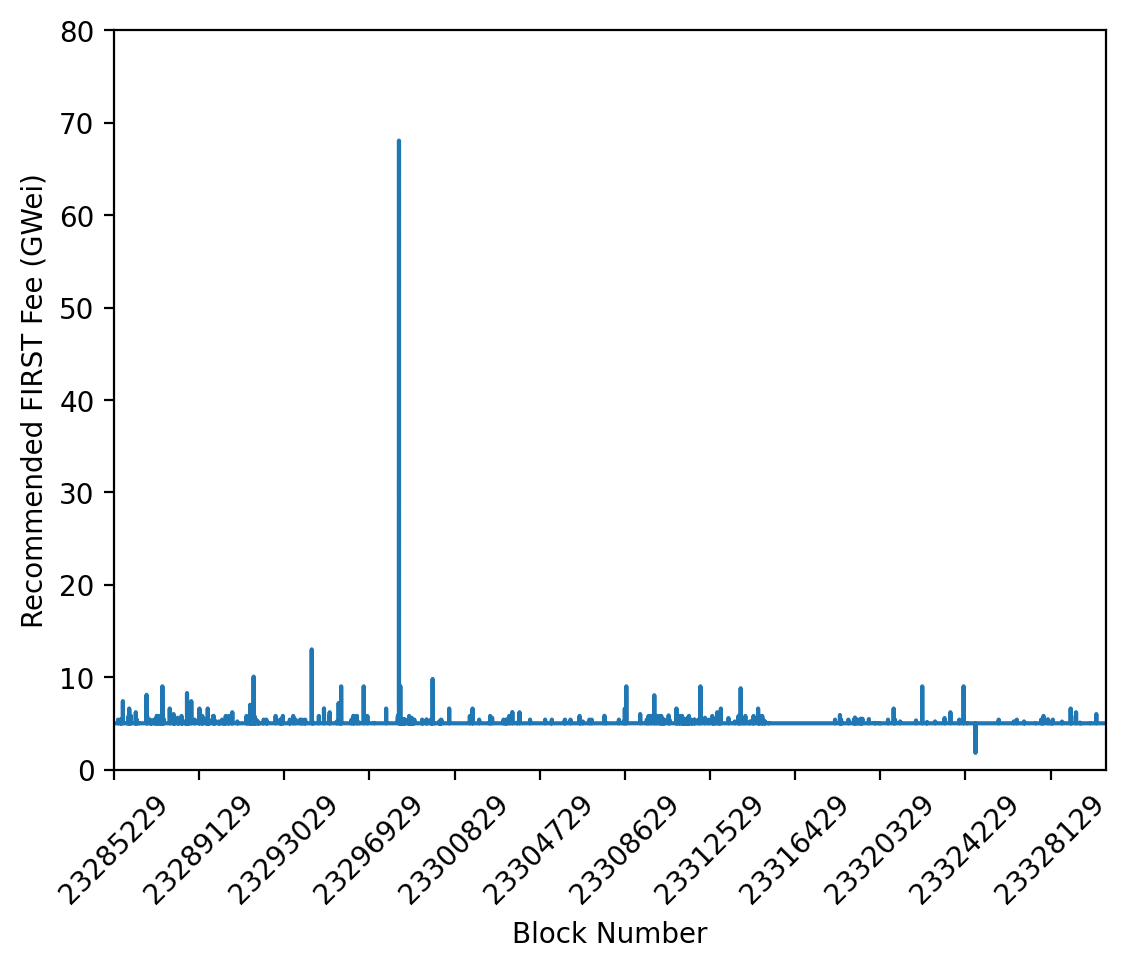}}\quad 
 
  \caption{Recommended FIRST fee for Ethereum and BSC blockchains per block. }
 \Description{}
\end{figure*}

We plot Figure~\ref{FirstFEEa} and Figure~\ref{FirstFEEb} to demonstrate how \FIRST\ recommended fee changed over our observation period in Ethereum and BSC blockchains, respectively. The figures show the recommended fee ($\firsttip{}$) on the Y-axes for the corresponding block number on the X-axes, computed using Protocol~\ref{proto:firstCalc}. In both experiments, $k=3$ and $\alpha=0.6$. For Ethereum $t_2$ was 30s and for BSC it was 5s. The $\firsttip$ calculated on Ethereum refers to the recommended priority fee, while on BSC, it refers to the recommended gas price.

\noindent In Figure~\ref{FirstFEEa}, the X-axis represents the 198K blocks on the Ethereum blockchain. As seen from the graph, the highest spike in our recommended fee is around block number 15697567. 
Some blocks have an associated spike in the recommended transaction $\firsttip$ due to the surge in the priority fees paid by transactions in the prior blocks. For example, the sale of tokens for the popular NFT project Art Blocks was confirmed in block number 15697567.  Out of the 446 transactions in this block, 405 purchased tokens using the Art Blocks contract and paid much higher priority fee than other network transactions. This affected the $\firsttip$ for 15697568. Similarly, the second-highest spike around block number 15741444 was due to the NFT project ``BeVEE - Summer Collection'' sales. 

The X-axis in Figure~\ref{FirstFEEb} represents the 41K blocks on the BSC blockchain. We see a spike in the FIRST fee for block 23298282 because four transactions indexed in the first four spots of the block 23298281 paid an average of 858.34 GWei in gas fee---escalating the recommended FIRST fee. On analyzing the block, we believe that the transactions paid high fee to profit from arbitrage opportunity. \change{Despite the unpredictable events in the Blockchain, Figures~\ref{FirstFEEa} and ~\ref{FirstFEEb} show that the computed $\firstfee$ adjusts to network activities.} In general, we noticed significantly less number of spikes in BSC, compared to Ethereum. This is due to the fast confirmation of transactions in BSC--more discussion at the end of this section.

To initiate our experiments, we obtained the $50^{th}$ percentile of the maximum wait time for the first 100 blocks and to better handle system dynamism, set $t_2$ to twice the value, $t_2=30$secs.
We also analyzed the Ethereum data for $\alpha =\{0.1,0.2,0.4,0.6,0.8\}$ and found that the $\alpha=0.6$ gives us better success rate than other values. Note that despite the occasional spikes most transactions pay a low priority fee, hence the value of $\alpha$ has limited impact. 

For our analysis, we set $k=3$, resulting in the VDF delay $t_1=90$secs. To reiterate our use cases discussion (Section~\ref{sec:limitation}), the VDF delay value is a function of the application and its risk appetite and can be tuned in \first{}. Even with $t_1=30$secs, only $0.004\%$ of transactions were susceptible to frontrunning!
We discuss this below.
Let $tx_{i}$ represent the $i^{th}$ transaction in a block ($b$), where $tx_{i}.fee$ and $tx_i.ctime$ are the transaction fee and the duration $tx_{i}$ waited on the mempool respectively, and $T_b$ represents the number of transactions in block $b$. Then, the fraction of potentially frontrunnable transactions in $b$ is given by,
  \begin{equation} 
  fr = \frac{
        \sum_{i=0}^{T_{b}}\llbracket tx_{i}.fee \geq \firsttip \rrbracket \llbracket tx_i.ctime \geq t_1\rrbracket}{T_{b}},
 \end{equation}
    where 
    $\llbracket.\rrbracket$ indicates \emph{Iverson brackets} such that $\llbracket i_{fee} \geq tip_{e}\rrbracket$ is {\em true} (1) if $i_{fee} \geq tip_{e}$, is {\em false} (0), otherwise. 
%

%

\begin{figure}[b]
 \centering
    \includegraphics[width=0.47\textwidth]{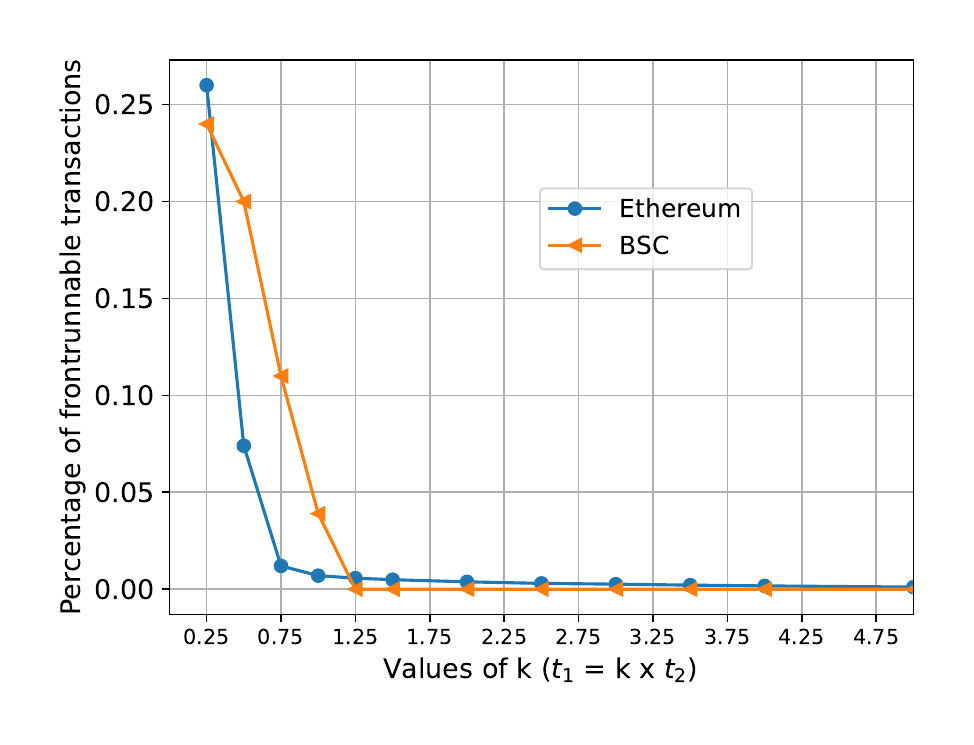}
    \vspace{-0.1in}
     \caption{Percentage of frontrunnable transactions (Y-axis) for different values of \first{} parameter $k$ ($t_1 = k\times t_2$).}
     \label{fig:differentT1s}
     \vspace{-0.2in}
\end{figure}

We analyzed the Ethereum and BSC data for different values of $k$. Figure~\ref{fig:differentT1s} shows the percentage ($fr \times 100$) of transactions that are frontrunnable out of the total transactions (24.34M in Ethereum and 5.14M in BSC) for different values of $k$. 
With the VDF delay of $90$s ($k=3$) and the FIRST recommended fee, on the Ethereum blockchain, 196319 out of 198235 blocks ($>99\%$) had no frontrunnable transactions! 
With $t_1=15$secs ($k=3$) and the FIRST recommended fee per BSC block, in BSC none of the transactions were frontrunnable.
In fact, the percentage of frontrunnable transactions goes to zero for $k \geq 2$.
Our choice of $k=3$ for the data is a good balance between the success rate and the imposed transactions delay. 

As we discussed before, on Ethereum, the chance of transactions being frontrun is a bit higher on account of higher volatility (we theorize, due to NFT transactions and slower block confirmation time) compared to BSC, which is more stable on account of the faster settling of transactions. For example, from our data, in the time it takes Ethereum to confirm one block, BSC confirms on an average 4.4 blocks. Each Ethereum block in our dataset has on an average 151 transactions, whereas it is 120 transactions in each BSC block. Thus, 666 BSC transactions are confirmed in the same time as 151 Ethereum transactions.

\section{Design Choices, Compatibility, Use Cases and Limitations}\label{sec:limitation}

In this section, we explore potential alternative solutions and their disadvantages, the design choices of \first{}, its compatibility with other protocols, and the limitations of our work.

\noindent \textbf{Verifiers}: One might question the need for a VDF in the presence of an honest-majority committee, which can be used to verify delays. 
A verifier-based approach would require each verifier to maintain a timer per request, which becomes infeasible as the transaction volume scales. Moreover, such delays cannot be independently verified. In contrast, VDFs are publicly verifiable and efficient to check.

\noindent  \textbf{Compatibility with private transaction}: The \first{} framework is designed to protect the transaction from getting frontrun. Since it does not change transaction structure, it is compatible with private relayers, such as Flashbots~\cite{flashbots}. The only requirement for 
a transaction before its submission to the relayers is to include the aggregated signature of $\verifiers$ on the verification of $\VDF$ proof (Protocol~\ref{proto:proofVerification}, Line 20). The $SC$ will assert if the transaction includes the aggregated signature and rejects it if not present. 
\FIRST{} independently prevents frontrunning attacks on EVM-based blockchains without needing extra protocols. While compatible with Flashbots, combining them is redundant and could compromise security through relayer delays.

\noindent \textbf{Potential use cases:} Ethereum Name Service (ENS)~\cite{ens} aims to map long and hard-to-memorize Ethereum addresses to human-readable identifiers. Recent sale trends and exorbitant offers, such as \emph{amazon.eth}, which received a million-dollar offer~\cite{amazonOffer}, indicate the importance of frontrunning prevention solutions. \first{} can be used during the sale of these domain names to prevent frontrunning. Non-fungible tokens (NFTs) are unique cryptographic tokens that live on blockchains and are not possible to forge. One of the largest NFT marketplace Opensea exceeded 10 billion dollars in NFT sales in the third quarter of 2021~\cite{yahoo}. Not surprisingly, frontrunning bots are watching the mempool for NFT sales to create a counter transaction to frontrun. One can employ \first{} to prevent such attacks on the marketplaces.

\noindent \revision{ \textbf{Compatibility with Time-Sensitive Applications:} 
\FIRST{} can be integrated with time-sensitive dApps, such as DEXes and NFT marketplaces, to prevent frontrunning attacks. When a platform adopts \FIRST{}, all user transactions experience a uniform VDF delay, ensuring that transactions are properly time-shifted and fairly ordered.
The VDF delay can be tuned to balance responsiveness and security, minimizing user inconvenience while maintaining strong protection guarantees.
In the event of a price discrepancy between a \FIRST{}-protected DEX and other exchanges, arbitrageurs will naturally intervene to close the price gap, ensuring consistency across platforms without undermining the protocol’s security.}

\noindent \textbf{Limitations:} Adjusting the real-world delay time with the given VDF delay parameter for every user's computational capabilities is a challenging and open-research problem~\cite{baum2021sok}. While it is an orthogonal task to ours, \first{} mitigates the problem by picking the $t_1 >> t_2$ --- this ensures that a more-capable Mallory cannot frontrun a less-capable Alice. \revision{The VDF delay parameter is selected based on observed transaction settlement times onchain and the computational capabilities of high-end contemporary machines. While not bulletproof, our analysis shows that over 99\% of transactions are no longer susceptible to frontrunning under \FIRST{}.}

Another limitation arises when an entity tries to re-submit a pending transaction created to interact with the \FIRST{} protected protocol, perhaps with a higher gas fee. Since the transaction is seen in the pending pool by all the entities, it increases the chances of getting frontrun. Lastly, our framework does not support the interaction of two \FIRST{} protected contracts, which we aim to address in future work. We note that \FIRST{} is a probabilistic solution as it recommends a fee to be paid by the users in the system to avoid getting frontrun with a high probability. However, as specified by the advantage statement in our theoretical analysis, there is a chance that a sufficiently funded and powerful adversary can outpace and frontrun honest users.  
To successfully frontrun a target user, an adversary not only needs commensurately larger computational resources than the norm to compute the VDF proof faster, but the adversary also needs to delay the target user's transaction in the mempool for the duration of time it takes to compute a valid VDF proof by inserting other transactions with higher fees than the target. This is a high barrier even for a very resourceful adversary. 
This is the price we pay for having an autonomous distributed system with no central control. Achieving zero frontrunning probability would require centralized transaction serialization, compromising blockchain decentralization and scalability.
\enlargethispage{-\baselineskip}

\section{Related Work}
\label{relatedworknew}
Current frontrunning research focuses on attack classification, profitability analysis, and mitigation strategies. We describe each below.

\noindent \textbf{Frontrunning prevention strategies}:  
Research in frontrunning prevention falls into three broad categories: (a) solutions that require direct interaction with miners to include the transaction in the upcoming block (private relayer approaches)~\cite{flashbots}; (b) solutions designed for DEXs (protocol incentive design)~\cite{zhou2020high, zhou2021a2mm}; and (c) solutions that prevent arbitrary reordering of transactions (order fairness)~\cite{kelkar2021order,kelkar2021themis,kursawe2020wendy}. In the first category, Alice sends $tx_{A}$ directly to miners via hidden endpoints to prevent adversaries from identifying her transaction in the pending pool. Flashbots~\cite{flashbots} is one example, where entities called \emph{relayers} bundle and forward transactions to miners through private channels. However, relayers themselves could perform frontrunning attacks as they have access to the complete transaction details. The second category of solutions is built for AMM-based DEXs; it reduces the risk of frontrunning by computing an optimal threshold for the frontrunner's transaction and routing the victim's swap request to minimize potential profit extraction. However, these solutions are specific to DEXs and cannot be applied to other dApps such as auctions, naming services, or games~\cite{eskandari2019sok, zhou2021a2mm, khalil2019tex}. The third category of solutions is built upon the order-fairness property, which ensures that the order of transactions in the finalized block reflects the order in which users submitted them~\cite{kelkar2021themis,kursawe2020wendy,kelkar2021order}. These solutions require significant changes to the consensus layer, making them impractical, while our approach works with existing EVM-based blockchains without modification.

\noindent \textbf{Surveys of frontrunning and related mechanisms}:
Eskandari \emph{et al.} presented a taxonomy of frontrunning attacks and analyzed the attack surface of top dApps~\cite{eskandari2019sok}. Qin \emph{et al.}~\cite{qin2021quantifying} extended the taxonomy of~\cite{eskandari2019sok} and quantified the profit made by blockchain extractable value~\cite{qin2021quantifying}. Daian \emph{et al.}~\cite{daian2019flash} revealed frontrunning bots competing in priority gas auctions and coined "Miner Extractable Value" (MEV) to describe miners reordering transactions for profit. \cite{baum2021sok} presented the state-of-the-art in frontrunning research and proposed a categorization of mitigation strategies. Additionally, Yang \emph{et al.} developed a taxonomy of MEV prevention solutions and conducted a comparative analysis of these approaches.

\noindent \textbf{Profitability analysis}: The profits made by frontrunners have been quantified by Torres \emph{et al.}~\cite{torres2021frontrunner} and Qin \emph{et al.}~\cite{qin2021quantifying}; the latter also brought to attention the presence of private transactions submitted to miners. Zhou \emph{et al.}~\cite{zhou2020high} formalized and quantified the profit made by \emph{sandwich attacks} enabled by frontrunning on decentralized exchanges. Qin \emph{et al.}~\cite{qin2020attacking} analytically evaluated Ethereum transactions' atomicity, analyzed two flash loan-based attacks, and demonstrated how attackers could have maximized their profit. Wang \emph{et al.}~\cite{wang2022cyclic} proposed a framework that analyzes the profitability conditions on cyclic arbitrage in DEXs. 

There are a couple of prior works~\cite{breidenbach2018enter,varun2022mitigating} that do not fall into any of the aforementioned three categories. LibSubmarine uses a commit-and-reveal scheme to prevent frontrunning~\cite{breidenbach2018enter}, where the committer must create a new smart contract for every transaction they submit to a dApp, which is inefficient. In a recent work, Varun \emph{et al.}~\cite{varun2022mitigating} proposed a machine learning approach to detect transactions that were frontrun in real-time. This approach requires the machine learning model to learn regularly. Further, the approach does not take into account priority fee, hence could fail to identify high priority fee based frontrunning transactions. There are also works that are related to our proposed scheme, such as Slowswap~\cite{slowswap}, which utilizes VDFs to introduce delays for transactions related to AMMs only. However, the current implementation employs a uniform VDF delay for all transactions, which is not ideal given the dynamic nature of Ethereum. In contrast, FIRST conducts statistical analysis and assigns VDF delay based on the network usage. Another solution in the MEV mitigation space is Radius~\cite{theradius}, which also aims to prevent frontrunning and sandwich attacks by implementing encrypted mempools. The Radius solution requires a mempool redesign, limiting its applicability, whereas \first{} integrates seamlessly with any EVM-based blockchain. 

\section{Conclusion} \label{conclusion}
We introduced \first{}, a decentralized framework aimed at mitigating frontrunning attacks on EVM-based smart contracts without necessitating changes to the blockchain consensus layer. Unlike application-specific approaches, \first{} is designed as a versatile and general-purpose solution, ensuring broad applicability across diverse dApps. Experimental results show that \first{} effectively reduces the likelihood of frontrunning attacks on two prominent blockchains: Ethereum and Binance Smart Chain. Additionally, the security guarantees of \first{} are rigorously established through the UC framework. \revision{In the future, we will explore gas golfing and better aggregate signature design to help reduce the gas fees needed for on-chain signature verification.}




\section{Acknowledgments}
This material is based upon work supported by the National Science Foundation under Award No 2148358, 1914635, 2417062, and the Department of Energy under Award No. DESC0023392. Any opinions, findings and conclusions, or recommendations expressed in this material are those of the authors and do not necessarily reflect the views of the National Science Foundation and the Department of Energy.

\bibliographystyle{plain}
\bibliography{references}

\begin{thebibliography}{10}

\bibitem{dydx}
d{Y}d{X}, 2024-7-7.
\newblock \url{https://dydx.exchange/}.

\bibitem{DefiLama}
{DeFi Lama}, 2024-12-08.
\newblock \url{https://defillama.com/chain/Ethereum}.

\bibitem{aave}
{AAVE}, 2024-7-7.
\newblock \url{https://aave.com/}.

\bibitem{ens}
{Ethereum Name Service}, 2024-8-12.
\newblock \url{https://ens.domains/}.

\bibitem{amazonOffer}
{Amazon.eth ENS domain owner disregards 1M USDC buyout offer on OpenSea}, 2024-8-12.
\newblock \url{https://cointelegraph.com/}.

\bibitem{yahoo}
{More than \$10bn in volume has now been traded on OpenSea in 2021}, 2024-8-12.
\newblock \url{https://yahoo.com}.

\bibitem{slowswap}
slowswap, 2023-8-12.

\bibitem{theradius}
theradius, 2023-8-12.

\bibitem{etherscanChart}
{Etherscan}, 2024-7-9.
\newblock \url{https://etherscan.io/chart/tx }.

\bibitem{yearn}
{yearn.finance}, 2024-7-7.
\newblock \url{https://yearn.finance/}.

\bibitem{compound}
{Compound Finance}, 2024-8-12.
\newblock \url{https://compound.finance/}.

\bibitem{baum2021sok}
Carsten Baum, James Hsin-yu Chiang, Bernardo David, Tore~Kasper Frederiksen, and Lorenzo Gentile.
\newblock Sok: Mitigation of front-running in decentralized finance.
\newblock In {\em International Conference on Financial Cryptography and Data Security}, pages 250--271. Springer, 2022.

\bibitem{boneh2018verifiable}
Dan Boneh, Joseph Bonneau, Benedikt B{\"u}nz, and Ben Fisch.
\newblock {Verifiable delay functions}.
\newblock In {\em Annual International Cryptology Conference}. Springer, 2018.

\bibitem{boneh2018survey}
Dan Boneh, Benedikt B{\"u}nz, and Ben Fisch.
\newblock {A Survey of Two Verifiable Delay Functions}.
\newblock {\em IACR Cryptol. ePrint Arch.}, 2018.

\bibitem{boneh2003aggregate}
Dan Boneh, Craig Gentry, Ben Lynn, and Hovav Shacham.
\newblock {Aggregate and verifiably encrypted signatures from bilinear maps}.
\newblock In {\em International conference on the theory and applications of cryptographic techniques}. Springer, 2003.

\bibitem{breidenbach2018enter}
Lorenz Breidenbach, Phil Daian, Florian Tram{\`e}r, and Ari Juels.
\newblock {Enter the Hydra: Towards Principled Bug Bounties and Exploit-Resistant Smart Contracts}.
\newblock In {\em 27th USENIX Security Symposium}, 2018.

\bibitem{canetti2004universally}
Ran Canetti.
\newblock {Universally composable signature, certification, and authentication}.
\newblock In {\em Proceedings. 17th IEEE Computer Security Foundations Workshop, 2004.}, 2004.

\bibitem{chaliasos2022study}
Stefanos Chaliasos, Arthur Gervais, and Benjamin Livshits.
\newblock A study of inline assembly in solidity smart contracts.
\newblock {\em Proceedings of the ACM on Programming Languages}, 6(OOPSLA2):1123--1149, 2022.

\bibitem{cohen2019chia}
Bram Cohen and Krzysztof Pietrzak.
\newblock {The chia network blockchain}, 2019.

\bibitem{daian2019flash}
Philip Daian, Steven Goldfeder, Tyler Kell, Yunqi Li, Xueyuan Zhao, Iddo Bentov, Lorenz Breidenbach, and Ari Juels.
\newblock {Flash Boys 2.0: Frontrunning in Decentralized Exchanges, Miner Extractable Value, and Consensus Instability}.
\newblock In {\em IEEE Symposium on Security and Privacy (SP)}, 2020.

\bibitem{cryptoeprint5}
Karim Eldefrawy, Sashidhar Jakkamsetti, Ben Terner, and Moti Yung.
\newblock Standard model time-lock puzzles: Defining security and constructing via composition.
\newblock Cryptology ePrint Archive, Paper 2023/439, 2023.
\newblock \url{https://eprint.iacr.org/2023/439}.

\bibitem{eskandari2019sok}
Shayan Eskandari, Mahsa Moosavi, and Jeremy Clark.
\newblock {Sok: Transparent dishonesty: front-running attacks on blockchain}.
\newblock {\em Financial Cryptography}, 2019.

\bibitem{kelkar2021order}
Mahimna Kelkar, Soubhik Deb, and Sreeram Kannan.
\newblock {Order-fair consensus in the permissionless setting}.
\newblock In {\em Proceedings of the 9th ACM on ASIA Public-Key Cryptography Workshop}, 2022.

\bibitem{kelkar2021themis}
Mahimna Kelkar, Soubhik Deb, Sishan Long, Ari Juels, and Sreeram Kannan.
\newblock Themis: Fast, strong order-fairness in byzantine consensus.
\newblock In {\em Proceedings of the 2023 ACM SIGSAC Conference on Computer and Communications Security}, pages 475--489, 2023.

\bibitem{khalil2019tex}
Rami Khalil, Arthur Gervais, and Guillaume Felley.
\newblock {TEX-A Securely Scalable Trustless Exchange.}
\newblock {\em IACR Cryptol. ePrint Arch.}, 2019.

\bibitem{kursawe2020wendy}
Klaus Kursawe.
\newblock {Wendy, the good little fairness widget: Achieving order fairness for blockchains}.
\newblock In {\em Proceedings of the 2nd ACM Conference on Advances in Financial Technologies}, 2020.

\bibitem{landerreche2020non}
Esteban Landerreche, Marc Stevens, and Christian Schaffner.
\newblock {Non-interactive cryptographic timestamping based on verifiable delay functions}.
\newblock In {\em International Conference on Financial Cryptography and Data Security}. Springer, 2020.

\bibitem{SEC}
Craig McCann.
\newblock {Detecting Personal Trading Abuses}, 2000.
\newblock \url{https://www.slcg.com/}.

\bibitem{bdos}
Michael Mirkin, Yan Ji, Jonathan Pang, Ariah Klages{-}Mundt, Ittay Eyal, and Ari Juels.
\newblock {BDoS: Blockchain Denial-of-Service}.
\newblock In Jay Ligatti, Xinming Ou, Jonathan Katz, and Giovanni Vigna, editors, {\em {ACM} {SIGSAC} Conference on Computer and Communications Security}, 2020.

\bibitem{flashbots}
Alex Obadia.
\newblock {Flashbots: Frontrunning the MEV Crisis}, 2024-7-7.
\newblock \url{https://medium.com/flashbots/frontrunning-the-mev-crisis-40629a613752}.

\bibitem{pietrzak2018simple}
Krzysztof Pietrzak.
\newblock {Simple verifiable delay functions}.
\newblock In {\em Innovations in theoretical computer science conference (ITCS)}. Schloss Dagstuhl-Leibniz-Zentrum fuer Informatik, 2018.

\bibitem{qin2021quantifying}
Kaihua Qin, Liyi Zhou, and Arthur Gervais.
\newblock {Quantifying blockchain extractable value: How dark is the forest?}
\newblock In {\em 2022 IEEE Symposium on Security and Privacy (SP)}. IEEE, 2022.

\bibitem{qin2020attacking}
Kaihua Qin, Liyi Zhou, Benjamin Livshits, and Arthur Gervais.
\newblock {Attacking the defi ecosystem with flash loans for fun and profit}.
\newblock In {\em International Conference on Financial Cryptography and Data Security}. Springer, 2021.

\bibitem{rivest1996time}
Ronald~L Rivest, Adi Shamir, and David~A Wagner.
\newblock {Time-lock puzzles and timed-release crypto}.
\newblock {\em Massachusetts Institute of Technology. Laboratory for Computer Science}, 1996.

\bibitem{solidproof}
solidproof.
\newblock solidproof, 2024-7-7.
\newblock \url{https://solidproof.io/kyc}.

\bibitem{EIGENLAYER}
EigenLayer Team.
\newblock Eigenlayer: The restaking collective, 2024-7-7.
\newblock \url{https://docs.eigenlayer.xyz/eigenlayer/overview/whitepaper}.

\bibitem{torres2021frontrunner}
Christof~Ferreira Torres, Ramiro Camino, et~al.
\newblock {Frontrunner jones and the raiders of the dark forest: An empirical study of frontrunning on the {Ethereum} blockchain}.
\newblock In {\em 30th USENIX Security Symposium}, 2021.

\bibitem{UNISWAP}
Uniswap.
\newblock Uniswap, 2024-8-12.
\newblock \url{https://uniswap.org/}.

\bibitem{varun2022mitigating}
Maddipati Varun, Balaji Palanisamy, and Shamik Sural.
\newblock {Mitigating Frontrunning Attacks in {Ethereum}}.
\newblock In {\em Proceedings of the Fourth ACM International Symposium on Blockchain and Secure Critical Infrastructure}, 2022.

\bibitem{eip1559}
vbuterin.
\newblock {EIP 1559 FAQ}, 2024-7-7.
\newblock \url{https://notes.ethereum.org/@vbuterin/eip-1559-faq}.

\bibitem{wang2022cyclic}
Ye~Wang, Yan Chen, Haotian Wu, Liyi Zhou, Shuiguang Deng, and Roger Wattenhofer.
\newblock {Cyclic Arbitrage in Decentralized Exchanges}.
\newblock {\em Available at SSRN 3834535}, 2022.

\bibitem{wesolowski2019efficient}
Benjamin Wesolowski.
\newblock {Efficient verifiable delay functions}.
\newblock In {\em Annual International Conference on the Theory and Applications of Cryptographic Techniques}. Springer, 2019.

\bibitem{zhou2021a2mm}
Liyi Zhou, Kaihua Qin, and Arthur Gervais.
\newblock {A2MM: Mitigating Frontrunning, Transaction Reordering and Consensus Instability in Decentralized Exchanges}.
\newblock {\em arXiv preprint arXiv:2106.07371}, 2021.

\bibitem{zhou2020high}
Liyi Zhou, Kaihua Qin, Christof~Ferreira Torres, Duc~V Le, and Arthur Gervais.
\newblock High-frequency trading on decentralized on-chain exchanges.
\newblock In {\em {2021 IEEE Symposium on Security and Privacy (SP)}}, 2021.

\end{thebibliography}

\appendix
\section{Definitions of Cryptographic Primitives}
\label{app:VDF}
\begin{definition}
(Verifiable Delay Function~\cite{boneh2018verifiable})  %
\label{VDF-appendix}
A verifiable delay function, $\VDF$
is defined over three polynomial time algorithms. 
\textbf(a) $\setup(\lambda,T) \rightarrow pp = (ek, vk)$: This is a randomized algorithm that takes a security parameter $\lambda$ and a desired puzzle difficulty $T$ and produces public parameters $pp$ that consists of an evaluation
key $ek$ and a verification key $vk$. We require $\setup$ to be polynomial-time in $\lambda$. By convention, the public parameters specify an input space $X$ and an output space $Y$. We assume
that $X$ is efficiently sampleable. $\setup$ might need secret randomness, leading to a scheme requiring a trusted setup. For meaningful security, the puzzle difficulty $T$ is restricted to be
sub-exponentially sized in $\lambda$.
\textbf(b) $\eval(ek, x)\rightarrow (y, \pi)$: This algorithm takes an input $x\in X$ and produces an output $y \in Y$ and a (possibly
empty) proof $\pi$. $\eval$ may use random bits to generate the proof $\pi$ but not to compute $y$. For
all $pp$ generated by $\setup(\lambda,T)$ and all $x \in X$ , algorithm $\eval(ek, x)$ must run in parallel time $T$ with poly$(\log(T), \lambda)$ processors.

     \textbf(c) $\verify (vk, x, y, \pi) \rightarrow$ \{``\text{accept}'', ``\text{reject}''\}: This is a deterministic algorithm that takes an input, output and proof
and outputs accept or reject. The algorithm must run in total time polynomial in $\log T$ and $\lambda$. Notice that $\verify$ is much faster than $\eval$.

\end{definition}

\begin{definition}(Aggregate signature~\cite{boneh2003aggregate})
An aggregate signature scheme is defined over five polynomial time algorithms: ($\KeyGen$, $\sign$, $\verify$, $\aggsign$, $\aggverify$). Let $\mathbb{G}_1$ and $\mathbb{G}_2$ be two multiplicative cyclic groups of prime order $p$ generated by $g_1$ and $g_2$, respectively. Let $\mathbb{U}$ be the universe of users. 
\noindent $\KeyGen(1^\lambda) \rightarrow (x_i, v_i)$: Each user picks random $x_i \leftarrow  \set{Z}_{p} $ and does $v_{i} \leftarrow g_{2}^{x_i}$. 
The user's public key is $v_{i} \in \set{G}_{2} $  and secret key is $x_{i} \in  \set{Z}_{p}$.\\
\noindent $\sign(x_i, M_i) \rightarrow \sigma_i$: Each user $i \in \set{U}$, given their secret key $x_{i}$ and message of their choice $M_{i}$ computes hash $h_i \leftarrow H(M_i)$ and signs $\sigma_{i} \leftarrow h_i^{x_i}$ where $\sigma_{i} \in \mathbb{G}_1$ and $H:\{0,1\}^{*} \rightarrow \mathbb{G}_1$.\\ 
\noindent $\verify(v_i, M_i, \sigma_i) \rightarrow$ \{$\true$, $\false$\}: Given public key $v_i$ of user $i$, a message $M_{i}$ and $\sigma_{i}$, compute $h_i \leftarrow H(M_i)$ and return true if $e(\sigma_{i}, g_2) = e(h_i, v_i)$.\\ 
\noindent $\aggsign(M_1, \dots, M_n, \sigma_1, \dots, \sigma_n) \rightarrow \sigagg$: Given each user $i$'s signature $\sigma_{i}$ on a message of their choice $M_{i}$, compute 
$\sigma_{agg} \leftarrow \prod_{i = 1}^{n} \sigma_{i}$ where $n = |U|$. \\
\noindent $\aggverify(\sigagg, M_1,\dots, M_n, pk_1,\dots,pk_n) \rightarrow$ $\{\true, \false\}$: To verify aggregated signature $\sigma_{agg}$, given original messages $M_{i}$ along with the respective signing users' public keys $v_{i}$, check if:
\begin{enumerate}
    \item All messages $M_{i}$ are distinct, and;
    \item For each user $i \in U$, $e(\sigma_i, g_{2}) = \prod_{i = 1}^{n} e(h_{i}, v_{i})$ holds true where $h_i \leftarrow H(M_{i})$. 
\end{enumerate}
\end{definition}

\section {EIP 1559} 
\label{EIP1599Com}
\revision{The London hard fork to Ethereum introduces novel transaction pricing mechanisms to improve the predictability of gas prices even during dynamic periods~\cite{eip1559}.
Users are now required to pay a \emph{base fee}, which is a fee computed according to a formula that may increase or decrease per block depending on network utilization. Besides a base fee, a user is encouraged to pay a \emph{priority fee} to incentivize the validators to prioritize the user's transactions. The transactions that follow EIP-1559 are termed {\em Type-2} transactions. While Ethereum has adopted EIP-1559, it's worth noting that other prominent blockchain networks, like Binance Smart Chain, have not yet implemented this standard. Despite the differing approaches, Ethereum and Binance Smart Chain remain two of the most widely used blockchain platforms. In our evaluation, we leverage these platforms as references to evaluate the proposed framework and demonstrate its applicability.}

\section{UC functionalities}~\label{UCFUNC}


\begin{figure}[h!]
\begin{mdframed}
\begin{center}
\textbf{Functionality} $\F{setup}$ 
\end{center}

\textbf{Setup}: On receiving tuple $(\mathsf{setup},t_1,t_2,d,k,\lambda,\alpha,sid)$ from dApp creator \emph{dAC}, $\F{setup}$ verifies that $t_1 > t_2$, if not return $\bot$. 
$\F{setup}$ sets value of VDF delay to $t_1$ ($s$ in $\F{vdf}^{\gamma}$), 
locally stores variables $d$, $k$, $\lambda, \alpha$ and initializes $\first$ recommended fee $\firsttip = 0$. 

\textbf{KeyGen}: Upon receiving a request $(\KeyGen, uid,sid)$ from user $u$, $\F{setup}$ calls $\F{sig}$ with $(\KeyGen, uid)$. When $\F{sig}$ returns $(\verkey, uid, pk_{u})$, $\F{setup}$ records the pair $(u,pk_{u})$ in $\idtable$ and returns $(\verkey, uid, pk_{u})$ to the user and $\simu$.

\textbf{Sign}: When $\F{setup}$ receives a request $(\sign, uid, m,sid)$ from user $u$, it forwards the request to $\F{sig}$, who returns $\{(\Signature, uid, m ,\sigma), \bot\}$. If return value is not $\bot$, $\F{setup}$ stores $(m, \sigma, pk_{u}, 1)$ in $\atable$, where $pk_{u}$ is $u$'s verification key created and stored during key generation. $\F{setup}$ forwards $(\Signature, uid, m, \sigma)$ to $u$ and $\simu$, else returns $\bot$ to both.

\textbf{Verify}: When $\F{setup}$ receives $(\verify, uid, m, \sigma, pk^{\prime},sid)$ from user $u$, it forwards the request to $\F{sig}$, who returns $(\Verified, uid, m, f)$, $f \in \{0,1,\phi\}$. $\F{setup}$ records $(m, \sigma, pk^{\prime}, f)$ in $\atable$ and returns $(\Verified, uid, m, f)$ to both user and $\simu$. 
\textbf{Aggregate Signature}: Upon receiving ($\aggsign, M_{1},\dots M_{n}, pk_{1},\dots, pk_{n}, \sigma_{V_{1}}\dots \sigma_{V_{n}},sid$) from $\coordinator$, $\F{setup}$ checks if a tuple ($\sigagg, M_{1}\dots M_{n}$, $pk_{1}, \dots, \pk_{n}$) already exists in $\stable$, if so, it forwards $(\mathsf{Aggregated},\sigagg)$ to $\coordinator$ and $\simu$. Else, it checks if $n$ $>$ $\lvert \verifiers\rvert /2$, if not then $\bot$ is returned to $\coordinator$ and $\simu$. If previous check passed, $\F{setup}$ generates a string $\sigagg \sample \{0,1\}^\lambda$, adds ($\sigagg, M_{1},\dots M_{n}, pk_{1}, \dots, pk_{n}$) to table $\stable$, and forwards $(\mathsf{Aggregated},\sigagg)$ to $\coordinator$ and $\simu$.

\textbf{Aggregate Verify}: Upon receiving ($\mathsf{aggVer}, \sigagg$, $M_{1},\dots M_{n}$, $pk_{1}, \dots, pk_{n},sid$) from an entity, $\F{setup}$ checks if tuple ($\sigagg, M_{1},\dots M_{n}, pk_{1},\dots,pk_{n}$) exists in $\stable$. If yes, it forwards ``accept'', else forward ``reject'' to the calling entity and $\simu$.

\textbf{Hash Interface:} On receiving a message $(hash,m,sid)$ from a user $u$, $\F{setup}$ checks if tuple $(m,h)$ exists in $\hashtable$. If so, it returns $h$ and exits. If not then $\F{setup}$ creates $h \sample \{0,1\}^{\lambda}$, adds $(m,h)$ to $\hashtable$, and returns $h$ to $u$.

\textbf{Calculate \first{} Fee}: For every new block mined, $\F{setup}$ sends $\getData()$ request to $\F{bc}$. 
$\F{setup}$ then checks priority fee ($f_{i}$ where $i \in [1 \dots n]$) paid by $tx_1 \dots tx_n$ transactions in the latest block that waited less than $t_2$ time, and calculates the value of $f_{avg} = 1/n \times \sum f_{i}$. $\firsttip = \alpha \times f_{avg} + (1 - \alpha) \times \firsttip$.

\textbf{Return \first{} fee}: On receiving request ($returnFee,sid$) from user, $\F{setup}$ returns current value of $\firsttip$.
\end{mdframed}
\caption{Ideal functionality for system setup and signatures.}
\label{fig:fsetup}
\Description{}
\end{figure}

\subsection{Proof of Theorem~\ref{thm:uc}}
\label{app:proof}








We assume the existence of eight tables: $\utable$, $\atable$, $\ctable$, $\stable$, $\idtable$, $\sctable$, $\bctable$ and $\txpooltable$ that store the internal state of $\F{FIRST}$ and are accessible at any time by $\F{setup}$ (Figure~\ref{fig:fsetup}), $\F{bc}$ (Figure~\ref{fig:fbcp1}), and $\F{construct}$ (Figure~\ref{fig:fcons}), which are time-synchronized functionalities. The $\utable$ is used to store user transaction specific information, $\atable$ is used to store the signatures of verifiers and users, $\ctable$ keeps track of VDF-specific challenges issued to users, $\stable$ stores the aggregated signatures of verifiers, and $\idtable$ stores the identifiers and keys of users. The $\sctable$ stores the deployed smart contract address and code, $\bctable$ stores the generated transactions, and $\txpooltable$ stores the current transaction pool. We assume that $\F{setup}$'s $t_1$ and $t_2$ time period verification implicitly checks that $t_1$ and $t_2$ are in the same unit of time (i.e., both are in seconds, minutes, etc.). 

We note that $\F{bc}$ does not completely follow EIP-1559 because Ethereum, like other real-world protocols and systems, is constantly evolving, and as these systems change the ideal world would need to be constantly updated to model the real world accurately. $\F{bc}$ incorporates block size based on maximum fees per block and the block hash rate, and is still general enough to model even non-EIP-1559 blockchains similar to the real world $\first$ protocol which is applicable to multiple blockchain types.

To make the presentation clear, for each corruption case, through a complete run of the protocol, we discuss the two worlds separately, and show that $\env$'s view will be the same.
\noindent
\textbf{Part 1}: Let us first consider the system and parameter setup described in Protocols~\ref{proto:SystemSetup}, ~\ref{proto:ParamGen2}, and~\ref{proto:firstCalc} .
$\env$ initializes $\F{bc}$ with $(init, sid,\mathbb{P},\mathbb{H})$. 

\noindent  \textbf{1) Case 0: All verifiers are honest} 
    
    a) \textbf{Real-world}: In the real-world (Protocols ~\ref{proto:SystemSetup}, ~\ref{proto:ParamGen2}, ~\ref{proto:firstCalc}), the \dac\, generates a smart contract, deploys it on the BC, and initializes $\firsttip$ calculation. The $n$ verifiers will generate their keypairs, $(pk_{i}, sk_{i}), i \in [1..n]$. $\env$ sees the $SC$ and each verifier's $pk$. \dac\, will pick a $T$, and initialize the $\VDF$ class group with a negative prime $d$. 
    Note that since all verifiers are honest, $\env$ does not get to see their internal state, and secret keys. 
    The view of $\env$ will be $(SC, pp = (\mathbb{G}, T,d),$ $pk_{1},\dots,pk_{n},\lambda, k, t_1,t_2,\alpha, $FIRST\_FEE$)$, where $\lambda$ is the security parameter, and $k$ is the multiplying factor for $t_2$. 
     
    b) \textbf{Ideal-world}: $\simu$ picks a security parameter $\lambda$, 
    $(T, k) \sample \mathbb{Z}^+$, negative prime $d$, vdf delay $t_1$, target mempool wait time $t_2$, and EWMA parameter value $\alpha$, and sends $(\mathsf{setup},t_1,t_2, d, k, \alpha)$ to $\F{setup}$ to start the $\firsttip$ calculation. 
    $\simu$ calls $\F{bc}$ with $(sid, \text{deploy}, SC.id, code)$ (this step is implicit in all the following game hybrids). $\adv$ sends $\getData()$ to $\F{bc}$ to get a copy of $SC$ (also including all contents of the blockchain). $\simu$ makes $n$ calls to $\F{setup}$, $(\KeyGen, vid_{i})$, $i \in [1..n]$. $\F{setup}$ returns $(\verkey, vid_{i}, pk_{i})$ to $\simu$. $\simu$ generates a random $\mathbb{G}$ such that $\mathbb{G} = Cl(d)$, and sets $pp = (\mathbb{G}, T, d)$. $\simu$ call \textbf{Return \first{} fee} to get computed value of $\firsttip$.
    Thus the view of $\env$ is the same as the real-world. The view of $\env$ will be $(SC, pp = (\mathbb{G}, T, d), pk_{1},\dots,pk_{n},\lambda$, $k, t_1,t_2,\alpha, \firsttip)$.
    
    \noindent
    \textbf{2) Case 1: Some verifiers are corrupted}
        
        a) \textbf{Real-world}: Per our adversary model, less than half of the verifiers can be corrupted. \dac\, deploys the SC on the blockchain, initializes $\firsttip$ calculation, and all verifiers will generate their keypairs. In this case, $\env$ will have access to both $pk$ and $sk$ of a corrupted verifier. $\env$ will also have access to the corrupted verifiers' $D_{i}= U_{i} = \emptyset$. 
        Verifiers, corrupt or otherwise, have no role to play in Protocol~\ref{proto:firstCalc}.  
        \dac\, will deploy the $SC$ on the blockchain as before, and will generate $\mathbb{G} = Cl(d), T$. Let the set of corrupted verifiers be $\mathbb{V}^{\prime}$, such that $\mathbb{V}^\prime \subset \mathbb{V}$, and $|\mathbb{V}^{\prime}| < |\mathbb{V}|/2$. The view of $\env$ will be $(SC, pp$  $= (\mathbb{G},$ $ T,$ $ d),$ $\{pk_{i},$ $sk_{i},$ $ D_{i},$ $ U_{i}\}_{i \in \mathbb{V}^{\prime}}, \{pk_{j}\}_{j \in (\mathbb{V})},$
        $\lambda, k, t_1,t_2,\alpha, \firsttip$).
        
        b) \textbf{Ideal-world}: As in Case 0, $\simu$ simulates \dac's role and receives from $\F{setup}$ $(\mathsf{Init},T, d,\firsttip)$. 
        $\simu$ calls $\F{bc}$ with $(\text{deploy}, SC.id, code)$. $\env$ sends $\getData()$ to $\F{bc}$ to get a copy of $SC$ (also including all contents of the blockchain). For the honest verifiers, $\mathbb{V} - \mathbb{V}^\prime$, $\simu$ creates $pk \sample \{0,1\}^\lambda$. Corrupt verifiers in $\mathbb{V}^{\prime} \subset \mathbb{V}$ are handled by $\env$. 
        Following the same procedure as in Case 0's ideal world, $\simu$ generates a random $\mathbb{G}$ s.t., $\mathbb{G} = Cl(d)$ and outputs $(SC, pp = (\mathbb{G}, T, d), pk_{1},\dots,pk_{n}, k)$. The view of $\env$, taking into account the additional information $\env$ has from corrupted verifiers will be $(SC, pp = (\mathbb{G}, T, d),$ $\{pk_{i}, sk_{i}, D_{i}, U_{i}\}_{i \in \mathbb{V}^{\prime}}$, $\{pk_{j}\}_{j \in \mathbb{V}}, \lambda, k, t_1,t_2,\alpha,\firsttip)$, which is the same as the real-world. 

\noindent
\textbf{Part 2}: Now, let us consider Alice's setup as given in Protocol~\ref{proto:setup}.
   
   \noindent
    \textbf{1) Case 0: Alice and all verifiers are both honest} 
        
        a) \textbf{Real world}: Alice generates $M_A$, hashes it, signs the digest, $h$: $\sigma_{A} \leftarrow \sign(h,sk_{A})$, and sends $(h, \sigma_{A})$ to all members of $\verifiers$. $\env$'s view will be $\emptyset$ (since all verifiers are honest, it does not have access to their inputs).
        
        b) \textbf{Ideal world}: $\simu$ simulates Alice and will receive $(\text{req, aliceID}, h, \sigma_{A})$ from $\F{construct}$. $\simu$ does not take any further actions.
    
    \noindent
    \textbf{2) Case 1: Alice is honest, some verifiers are corrupt} 
        
        a) \textbf{Real world}: Alice generates the $(M_{A}, h, \sigma_{A})$ as in Case 0, and sends $(h, \sigma_{A})$ to $\verifiers$. If $\mathbb{V}^{\prime}$ is the set of corrupted verifiers, $\env$'s view will consist of $\mathbb{V}^{\prime}$'s inputs, i.e., $(\{sk_{i}\}_{i \in \mathbb{V}^{\prime}}, h, \sigma_{A})$. 
       
        b) \textbf{Ideal world}: $\simu$ generates an $h \sample \{0,1\}^k$ (note that verifiers do not know the preimage). $\simu$ then calls $\F{setup}$ with $(\sign, aid, h)$, where $aid$ is chosen at random. $\F{setup}$ returns $(\Signature,aid, h, \sigma_{aid})$. $\simu$ outputs $(h, \sigma_{aid})$. 
    
    \noindent
    \textbf{3) Case 2: Alice is corrupt and all verifiers are honest}
        
        a) \textbf{Real world}: Alice generates $M_A$ and $\sigma_{A}$ over $M_A$'s digest. If the signature does not verify, verifiers will eventually return $\bot$. If Alice does not send anything, verifiers will do nothing. In any case, $\env$'s view will be $(M_{A} = (addr_{A}, f_{\text{name}}, addr_{SC}), h, \sigma_{A})$.
        
        b) \textbf{Ideal world}: $\simu$ gets $(h, \sigma_{A})$ from $\env$. $\simu$ does not take any further actions. $\env$'s view will be $(M_{A} = (addr_{A}, f_{\text{name}}, addr_{SC}), h, \sigma_{A})$. 
    
    \noindent
    \textbf{4) Case 3: Both, Alice and some verifiers are corrupt} Note that this cannot be locally handled by $\env$, as one might expect, since some verifiers are still honest.
        
        a) \textbf{Real world}: Alice's actions will be the same as in Case 2's real world. $\env$'s view will be $(M_{A}, h, \sigma_{A}, \{sk_{i}\}_{i \in \mathbb{V}^{\prime}})$, where $\mathbb{V}^\prime$ is the set of corrupted verifiers.
        
        b) \textbf{Ideal world}: $\simu$ gets $(h, \sigma_{A})$ from $\env$. $\simu$ does not take any further actions. $\env$'s view is same as real world.

\noindent
\textbf{Part 3}: Now let us consider Alice, $\coordinator$, and verifiers' interaction as given in Protocol~\ref{proto:exchange}, ~\ref{proto:proofVerification}, and ~\ref{proto:smart:verify}. In the following cases, whenever some verifiers ($\verifpr$) are corrupt, $|\verifpr| < |\verifiers|/2 $, hence, a majority of verifiers are still honest.

    
    \noindent
    \textbf{1) Case 0: Alice, $\coordinator$ and all verifiers are honest} 
    
    a)  \textbf{Real-world}: On receiving a new VDF request from Alice, $\coordinator$ picks an $l \sample \mathbb{P}$, all verifiers send $(M_{i}, \sigma_{V_i})$ to $\coordinator$. $\coordinator$ will verify the signatures, and will return the aggregate signature $\sigagg$ to Alice, who will then compute the VDF proof, $(\pi, y)$. This proof is sent to the $\verifiers$ who will verify it before submitting their signatures to the $\coordinator$ for aggregation. The aggregated signature is sent to Alice by the $\coordinator$ who verifies it, and eventually submits $tx_{A}$ with the current $\firsttip$ to the $SC$ using $(sid,\mathsf{invoke},tx_{A})$. $\env$'s view will only be $\{pk_{i}\}_{i \in \mathbb{V}}$ initially, and it will see $tx_{A}$ only when it hits the transaction pool.

    b) \textbf{Ideal-world}: $\simu$ creates $M_A$, creates hash $h_A = H(M_A)$ and calls $\F{setup}$ and gets $\sigma_A$. It then forwards $(\mathsf{req},h_A,\sigma_A)$ to $\F{construct}$'s \textbf{User request} function and receives ``success" and $(\mathsf{req}, \text{aliceID}, h_A,\sigma_A)$. $\simu$ generates $l$ on behalf of the $\coordinator$ and sends $(l,\text{aliceID})$ to $\F{construct}$ using \textbf{Coordinator request} function and it receives $(\mathsf{valid}, l, h_A, \sigma_{A}, pk_{A})$. For each $\verifier_i$, $\simu$ signs the $M_i =  (l,h_A, \verifier_{i},block_{curr})$ using $\F{setup}$ and $block_{curr}$ is retrieved by calling $\F{bc}$, i.e., using $\getBlockNum()$ and being returned $ block_{curr} \leftarrow \lceil num_{tx}/block_{qty}\rceil$. $\simu$ sends $(\verifier_i, M_i,\sigma_{M_i})$ to $\F{construct}$ using \textbf{Coordinator response} function call. $\simu$ then simulates the aggregation step of $\coordinator$ by calling $\F{setup}$ and receives $\sigagg$. $\simu$ calls $\F{construct}$ \textbf{User response} to send $\sigagg$ to Alice. $\simu$ calls $\F{vdf}$ $(\text{start},l)$ function to start $\VDF$ delay. After $\VDF$ delay time, $\simu$ calls $(\text{output},l)$ function in $\F{vdf}$ to generate the $\VDF$ proof which returns $(s,p)$. $\simu$ then verifies the proof calling $(\text{verify},l,p,s)$ on behalf of each $\verifier_i$ and generates $M^{\prime}_i$ and $\sigma_{\verifier_i}^\prime$ in a straightforward way. $\simu$ aggregates all the signatures from $\verifiers$ using $\F{setup}$ and generates $\sigaggpr$ and forwarded to Alice using $\F{construct}$'s $\textbf{User response}$ function. $\simu$ creates $tx_A = (\sigma_{A}^{\prime}, M^{\prime}, pk_{A},\firsttip)$, where $M^\prime = (\messages_A, M_{\agg}, M_{\agg}^{\prime})$, $M_{\agg}^{\prime} = (\sigagg^{\prime}, M_{1}^{\prime},\dots,M_{n}^{\prime}, pk_{\verifier_1},\dots,pk_{\verifier_n})$, $\messages_{\agg} = (\sigagg,$ $M_1, \dots, M_n,$ $pk_{\verifier_1}$ $\dots pk_{\verifier_n})$, and $\firsttip$ is retrieved by sending ($returnFee$) request to $\F{setup}$. 
    $\simu$ sends ($sid, \mathsf{invoke},tx_A$) to $\F{bc}$ calling smart contract with $SC.id$. $tx_{A}$ will get added to the $\txpooltable$. While the $tx_{A}$ is in the pending pool, the adversary can try to delay $tx_{A}$ from being mined by submitting transactions with higher fees than $tx_{A}$, while the adversary generates a valid $\VDF$ proof. We point out that the adversary will need to submit enough transactions with higher fees so that $tx_{A}$ does not appear in any of the blocks before adversary has a valid $\VDF$ proof, which would require an exorbitant amount of fees just like in the real world. When $tx_{A}$ eventually gets mined and added to a block, it will appear in $\bctable$ with the corresponding $\blocknum$ and if an adversary's valid transaction did not get mined before Alice's then Alice did not get frontrunned (this step is implicit in all the following game hybrids). 
    The $code$ associated with $SC.id$ checks that the $\sigagg$ and $\sigaggpr$ are signed by majority $\verifiers$, $block_{curr}$ signed in $\sigagg$ is valid, and that the $h_A \overset{?}{=} H(M_A)$. If any of the checks fail, the smart contract returns $\bot$, else outputs a valid transaction $tx^\prime$. 
    
    
    
    \noindent
    \textbf{2) Case 1: Alice, $\coordinator$ are honest, some verifiers are corrupt}
    
    a)  \textbf{Real-world}: Honest $\coordinator$ generates $l \sample \mathbb{P}$. Corrupted verifiers can either: 1) deliberately fail Alice's signature verification (Step 4 of Protocol~\ref{proto:exchange}), or 2) create a bogus signature over a possibly incorrect message (Steps 5, 6 of Protocol~\ref{proto:exchange}). In both cases, the corrupt verifiers in $\mathbb{V}^\prime$ will not contribute towards $\sigagg$, since the $\coordinator$ needs a majority to abort the process and return $\bot$ (Step 18 of Protocol~\ref{proto:exchange}). As long as we have a honest majority in $\mathbb{V}$, honest $\coordinator$ will create and return $\sigagg$ to Alice, who will then evaluate the VDF and generate $(\pi,y)$, and send $(\pi, y)$ to all members of $\verifiers$. Similarly, during the generation of $\sigagg^\prime$, $\coordinator$ can ignore the inputs from $\verifiers^\prime$. $\coordinator$ sends $(\messages_\agg, block_{curr})$ to Alice. Honest Alice eventually outputs $tx_A$. $\env$'s view will be $(l, \sigma_{A}, h, pk_{A}, \pi, y, tx_{A}, block_{curr}, pp, t_1, t_2,\alpha)$. \footnote{We note that $\env$ always has access to all the $\BC$ data: In the real world $\env$ can query a full node, run a light node, etc. In the ideal world, $\env$ can send a $\getData()$ request to $\F{bc}$. Without loss of generality, we say the $\env$'s view includes $block_{curr}$ because a given $block_{curr}$ is only tied to the current request and signifies the current block number on the $\BC$ when the VDF request was received by the 
    verifiers.}

    b) \textbf{Ideal-world}: $\simu$ needs to simulate the actions of $\coordinator$ and Alice to $\env$. $\simu$ creates $M_A$, creates hash $h_A = H(M_A)$ and calls $\F{setup}$ and gets $\sigma_A$. It then forwards $(\mathsf{req},h_A,\sigma_A)$ to $\F{construct}$ and receives ``success'' and $(\mathsf{req}, \text{aliceID}, h_A,\sigma_A)$. $\simu$ picks an $l \sample \mathbb{P}$, calls \textbf{Coordinator request} function in $\F{construct}$, and sends $l$ to $\env$. If members of $\mathbb{V}^\prime$ return $\bot$ for Alice's signature verification or return bogus signatures from $\verifiers^{\prime}$ ($\simu$ can check these using \textbf{Verify} function call in $\F{setup}$), $\simu$ ignores them, since $|\mathbb{V}^{\prime}| < |\mathbb{V}|/2$. 
    $\simu$ then calls $\F{setup}$'s \textbf{Aggregate Signature} function with $(\aggsign,$ $M_{1},\dots,M_{n},$ $pk_{1},\dots,pk_{n},$ $\sigma_{V_{1}},\dots,\sigma_{V_n})$ to aggregate all honest majority $\verifiers$'s signatures. $\F{setup}$ returns $\sigagg$ to $\simu$. $\simu$ then calls $\F{vdf}$ to generate $\VDF$ proof $(s,p)$. $\simu$ sends $(l,p,s)$ to $\env$. Members of $\verifiers^\prime$ will send $\{\sigma^{\prime}_{i}\}_{i \in \verifiers^\prime}$ to $\simu$. $\simu$ will simulate signatures for members of ($\verifiers - \verifiers^\prime$) in a  straightforward way and calls $\F{bc}$'s $\getBlockNum()$ function to get the $block_{curr} \leftarrow \lceil num_{tx} /block_{qty} \rceil$ value. $\simu$ generates the $\sigagg^\prime$ similar to the previous $\sigagg$, by taking the majority signatures. Since members of $\verifiers^\prime$ will be in a minority, even if they return $\bot$, it will not affect the creation of $\sigagg^\prime$. Finally $\simu$ generates Alice's signature over $M^\prime$, creates $tx_{A}$ and submits it to $\F{bc}$. The view of $\env$, who controls $\verifiers^\prime$ will be $(l, \sigma_{A}, h_A, pk_{A}, p, s, tx_{A}, block_{curr}, pp,t_1,t_2,\alpha)$ 
    
    \noindent
    \textbf{3) Case 2: Alice, all  verifiers are honest, $\coordinator$ is corrupt}
    
    a) \textbf{Real-world}: On receiving VDF request from Alice, corrupt $\coordinator$ could either pick $l \sample \mathbb{P}$ which has already been assigned to another user or an $l \notin \mathbb{P}$, in this case the honest members of $\verifiers$ identifying the $\coordinator$ as corrupt, will not generate an accept message which can be aggregated by the $\coordinator$ and the $\coordinator$ cannot proceed. If the $\coordinator$ had picked $l \sample \mathbb{P}$ correctly, on receiving the accept messages and signatures from $\verifiers$, $\coordinator$ can still choose to create $\sigagg$ that would fail verification, in this case Alice's checks would fail, identifying the $\coordinator$ as corrupt and she would not proceed further with the protocol. If the $\coordinator$ had created $\sigagg$ correctly, Alice would generate the $\VDF$ proof and send it for verification to all $\verifiers$. Upon verification, $\verifiers$ send their replies to $\coordinator$ for aggregation. Like the previous aggregation step, if the $\coordinator$ creates a corrupt message in this step, Alice would be able to identify the $\coordinator$ as malicious. If the $\coordinator$ sends Alice a valid $\sigagg^{\prime}$, Alice eventually outputs $tx_A$. $\env$'s view will be $(l, \sigma_{A}, h, pk_{A}, \pi, y, tx_{A}, block_{curr}, pp$,$t_1,t_2,\alpha$ $,\sigma_{V_i}$,  $\messages_{i}$, $\sigma_{V^{\prime}_{i}}$, $M_{i}^{\prime})$ for $i \in [1 \dots | \verifiers | ]$. 

b) \textbf{Ideal-world}: $\simu$ needs to simulate the actions of $\verifiers$ and Alice to $\env$. $\env$ picks an $l \sample \mathbb{P}$, and sends to $\simu$.  $\simu$ sends $(l,\text{aliceID})$ to $\F{construct}$ and if it received $(\text{used},l)$ then $l$ has been used before and $\simu$ would return $\bot$ stopping the protocol. If $\env$ picked a valid $l$, $\simu$ simulates the operations of the honest $\verifiers$. $\simu$ sends each $V_i$'s accept message to $\env$ who creates an $\sigagg$ by calling $\F{setup}$'s Aggregate Signature function call. $\simu$ verifies $\sigagg$ before proceeding, else return $\bot$. This is sent to $\simu$ who simulates Alice's operation of computing the VDF, before simulating the members of $\verifiers$'s response accepting Alice's VDF proof computation, and forwarding $(M_{1}^{\prime},\dots M_{n}^{\prime}, pk_{1},\dots, pk_{n}, \sigma_{V_{1}}^{\prime}\dots \sigma_{V_{n}}^{\prime})$ to $\env$. If $\env$ does not aggregate the signatures from $\verifiers$ correctly, and sends corrupted/malformed $\sigaggpr$ to $\simu$, the signature verification by $\simu$ would fail. Finally $\simu$ simulates Alice's signature over $tx_{A}$ and submits to $\F{bc}$, i.e., $(sid, \mathsf{invoke}, tx_{A})$. The view of $\env$, who controls $\coordinator$ will be $(l, \sigma_{A}, h, pk_{A}, p, s, tx_{A}, block_{curr}, pp$, $t_1,t_2, \alpha$,  $\sigma_{V_i}$, $\messages_{i}$, $\sigma_{V^{\prime}_{i}}$, $M_{i}^{\prime})$ for $i \in [1 \dots | \verifiers | ]$.
    
    \noindent
    \textbf{4) Case 3: Alice is honest, $\coordinator$ and some verifiers are corrupt}

    a) \textbf{Real-world}: On receiving VDF request from Alice, corrupt $\coordinator$ could pick $l \sample \mathbb{P}$ which has already been assigned to another user or an $l \notin \mathbb{P}$, in this case the honest majority of $\verifiers - \verifpr$ would not generate a signature for $\coordinator$ identifying the $\coordinator$.  The corrupt $\verifpr$ could choose to generate accept messages and send them to $\coordinator$. The $\coordinator$ can create $\sigagg$ using the corrupt $\verifpr$'s accept messages but this would fail verification on when Alice receives $\sigagg$ as $| \verifpr | < | \verifiers | /2$. If the $\coordinator$ had picked $l \sample \mathbb{P}$ correctly, on receiving the accept messages and signatures from $\verifiers$, $\coordinator$ can still choose to create $\sigagg$ that would fail verification, in this case Alice's checks would fail, identifying the $\coordinator$ as corrupt and she would not proceed further with the protocol. If the $\coordinator$ had created $\sigagg$ correctly, Alice would generate the $\VDF$ proof and send it for verification to all $\verifiers$. Upon verification, honest members $\verifiers - \verifpr$, send their replies to $\coordinator$ for aggregation. The dishonest members $\verifpr$ could either choose to not send a valid $``accept''$ message for aggregation or choose to send a corrupt message for aggregation. The $\coordinator$ could choose to create a corrupt message by aggregating less than $| \verifiers | /2$ messages or create a junk $\sigaggpr$. Like the previous aggregation step, if the $\coordinator$ creates a corrupt $\sigaggpr$ in this step, Alice would be able to identify the $\coordinator$ as malicious because of the checks she does on receiving the messages from $\coordinator$. If the $\coordinator$ sends Alice a valid $\sigaggpr$, Alice eventually outputs $tx_A$. $\env$'s view will be $(l, \sigma_{A}, h, pk_{A}, \pi, y, tx_{A}, block_{curr}, pp, t_1,t_2,\alpha, \sigma_{V_i},$  $\messages_{i}$, $\sigma_{V^{\prime}_{i}}$, $M_{i}^{\prime})$ for $i \in [1 \dots | \verifiers | ]$.
    
    b) \textbf{Ideal-world}: $\simu$ needs to simulate the actions of $\verifiers - \verifpr$ and Alice to $\env$. $\env$ picks an $l \sample \mathbb{P}$, and sends to $\simu$. If $l$ has been used before, then $\simu$ would just return $\bot$ on behalf of honest $\verifiers$. $\env$ can still choose to create a $\sigagg$ with $\verifpr$ accept messages but when this is sent to $\simu$ it would fail verification since $| \verifpr | < | \verifiers | / 2$. If $\env$ picked a valid $l$, $\simu$ simulates operations of the honest verifiers and sends each $V_i \in \{\verifiers - \verifpr\}$ accept message to $\env$ who creates an $\sigagg$. This is sent to $\simu$ who computes the $\VDF$ proof and sends to $\env$. $\simu$ also send accept messages from honest $\verifiers$ to $\env$ for $\coordinator$'s operations. Like the previous aggregation step, $\env$ could choose to create corrupt $\sigaggpr$ but this would fail verification when sent to $\simu$ and the protocol would not proceed further. To proceed further, $\env$ has to create a valid $\sigaggpr$ with the accept messages from $> | \verifiers | /2$ members of $\verifiers$. Finally $\simu$ simulates Alice's signature over $(\sigagg^\prime, M_{A}, M_{1}^{\prime},\dots,M_{n}^{\prime}, pk_{1},\dots,pk_{n})$, creates $tx_{A}$ and submits to $\F{bc}$. The view of $\env$, who controls $\coordinator$ and $\verifpr$ will be $(l, \sigma_{A}, h, pk_{A}, p, s, tx_{A}, block_{curr}, pp, t_1,t_2,\alpha, \sigma_{V_i},$  $\messages_{i}$, $\sigma_{V^{\prime}_{i}}$, $M_{i}^{\prime})$ for $i \in [1 \dots | \verifiers | ]$.
    
    \noindent {\bf 5) Case 4: Alice is corrupt, $\coordinator$ and all verifiers are honest}
    
    a) \textbf{Real-world}: Alice sends a $\VDF$ request to $\coordinator$ and $\verifiers$. A corrupt Alice could choose to create a corrupt $\sigma_A$ but this would fail verification at the $\coordinator$ and $\verifiers$ and the protocol would stop. To proceed Alice has to compute valid $(\sigma_A, h)$. $\coordinator$ and $\verifiers$ would proceed as normal and return a $\sigagg$ to Alice. Alice could choose to send a corrupt $\VDF$ for verification to $\verifiers$. Since verification would fail there would be no $\sigaggpr$ generated for Alice so she cannot proceed further. If Alice computes a valid $\VDF$ proof, $\coordinator$ would return a $\sigaggpr$ to her and Alice eventually outputs $tx_A$. In $tx_A$ Alice could choose to use a different $M_A^{\prime}$ but the hash of $M_A^{\prime}$ would not match the $h$ signed in $\sigagg$ and would fail verification in the smart contract which checks $h$ matches $M_A$, and the $l$ and $h$ in $\sigaggpr$ are tied to $M_A$. Alice can only pass smart contract verification if she keeps the original $M_A$ and valid $\messages_{\agg}$ and $\messages_{\agg}^{\prime}$ in $tx_A$.  $\env$'s view will be $(l, \sigma_{A}, h, pk_{A}, \pi, y, tx_{A}, block_{curr}, pp, t_1,t_2,\alpha, \sigma_{V_i}$, $\messages_{i}$, $\sigma_{V^{\prime}_{i}}$, $M_{i}^{\prime})$ for $i \in [1 \dots | \verifiers | ]$.

    b) \textbf{Ideal-world}: $\simu$ needs to simulate the actions of $\verifiers$ and $\coordinator$ to $\env$. $\env$ picks a $M_A$ and sends a request to $\coordinator$ with hash $h$ corresponding to $M_A$. $\simu$ simulates $\coordinator$ and $\verifiers$ by assigning $l$ to $h$ and generating a $\sigagg$. $\sigagg$ is sent to $\env$ who computes the $\VDF$ proof. 
    If $\mathcal{Z}$  decides to send a corrupted proof to $\mathcal{S}$, 
    then it would fail verification and $\simu$ would not generate a corresponding $\sigaggpr$. 
    The only way for $\env$ to proceed is to compute valid $\VDF$ proof. 
    Upon receiving valid proof $\simu$ verifies it and generates $\sigaggpr$ which is sent to $\env$. $\env$ now creates $tx_A$ and submits to $\F{bc}$. The $code$ associated with $SC.id$ verifies $\sigaggpr$, the hash of $M_A$ included in $tx_A$ matches $(h,l,\cdot)$ in $\sigagg$, and the $l$ in previous tuple is same as in $\sigaggpr$. The code also checks for freshness using the $block_{curr}$ value. If any of these checks fail verification then the smart contract would not execute in favor of $\env$ and it would be identified as corrupt. $\env$'s view will be $(l, \sigma_{A}, h, pk_{A}, p, s, tx_{A}, block_{curr}$, $pp$, $t_1, t_2, \alpha$, $\sigma_{V_i}$, $\messages_{i}$, $\sigma_{V^{\prime}_{i}}$, $M_{i}^{\prime})$ for $i \in [1 \dots | \verifiers | ]$. 

        
        \noindent
        \textbf{6) Case 5: Alice and some verifiers are corrupt, $\coordinator$ is honest}
    
    a) \textbf{Real-world}: As in Case 4, if Alice sends corrupt $\sigma_A$ the $\coordinator$ would fail verification and not proceed further. If the $\sigma_A$ is valid, the $\coordinator$ picks valid $l$ and sends to all $\verifiers$. The corrupt minority of $\verifpr$ could choose to not send their signatures or send corrupt signatures which the $\coordinator$ can discard and generate a $\sigagg$ from the honest majority in $\verifiers$. Alice on receiving the $\sigagg$ can choose to send an invalid $\VDF$ proof which would not generate accept signatures from the honest majority in $\verifiers$. $\verifpr$ could choose to wrongly send accept signatures to $\coordinator$ but since $| \verifpr | < | \verifiers |/2$,  $\coordinator$ will not generate a $\sigaggpr$. If Alice computed a valid $\VDF$ proof then she will receive a $\sigaggpr$ from $\coordinator$ and Alice eventually outputs $tx_A$. As described in Case 4, Alice can only pass smart contract verification if she outputs a valid $tx_A$. $\env$'s view will be $(l, \sigma_{A}, h, pk_{A}, \pi, y, tx_{A}, block_{curr},$ $pp$, $t_1, t_2, \alpha$, $\sigma_{V_i}$,  $\messages_{i}$, $\sigma_{V^{\prime}_{i}}$, $M_{i}^{\prime})$ for $i \in [1 \dots | \verifiers | ]$.
    
    b) \textbf{Ideal-world}: $\simu$ needs to simulate the actions of $\verifiers - \verifpr$ and $\coordinator$ to $\env$. $\env$ picks a $M_A$ and sends a request to $\coordinator$ with hash $h$ corresponding to $M_A$. If $h$ or $\sigma_A$ are invalid then $\simu$ would not generate $l$ and the protocol would stop. If valid request is received from $\env$, $\simu$ assigns $l$ to $h$ and sends to $\env$. If $\verifpr$ controlled by $\env$ send corrupt signatures to $\simu$, it can just ignore those messages and output a $\sigagg$ to $\env$ by simulating the honest majority of $\verifiers$'s actions.  
    If $\env$ decides to send corrupt proof to $\simu$, then it would fail verification and $\simu$ would not generate a corresponding $\sigaggpr$. As in previous step, corrupt $\verifpr$ messages corresponding to $\VDF$ proof from $\env$ can be ignored by $\simu$. The only way for $\env$ to proceed is to compute valid $\VDF$ proof. Upon receiving valid proof $\simu$ verifies it and generates $\sigaggpr$ which is sent to $\env$. $\env$ now creates $tx_A$ and submits to $\F{bc}$.  As described in Case 4, $\env$ can only pass smart contract verification if  $tx_A$ contains valid signatures and $M_A$. $\env$'s view will be $(l, \sigma_{A}, h, pk_{A}, p, s, tx_{A}, block_{curr}$, $pp$, $t_1, t_2, \alpha$, $\sigma_{V_i}$,  $\messages_{i}$, $\sigma_{V^{\prime}_{i}}$, $M_{i}^{\prime})$ for $i \in [1 \dots | \verifiers | ]$.

        
        \noindent
        \textbf{7) Case 6: Alice, $\coordinator$ are corrupt, all verifiers are honest}
    
    a) \textbf{Real-world}: If Alice sends corrupt $\sigma_A$ to the $\coordinator$ and $\verifiers$, or if Alice sends a valid request but $\coordinator$ chose a corrupt $l$ similar to Case 2, the $\verifiers$ will not send accept signatures to $\coordinator$ since the request or $l$ will not pass verification. The $\coordinator$ can create corrupt $\sigagg$ but this would fail verification eventually at the smart contract. If the $\sigma_A$ is valid and the $\coordinator$ picks valid $l$, the $\verifiers$ will reply with accept messages so $\coordinator$ can generate a valid $\sigagg$. 
    Alice on receiving the $\sigagg$ can choose to send an invalid $\VDF$ proof which would not generate accept signatures from the $\verifiers$. 
    Like the previous stage, the $\coordinator$ can create corrupt $\sigaggpr$ but this would fail verification eventually at the smart contract.
    If Alice computed a valid $\VDF$ proof, the $\coordinator$ would receive accept signatures from the $\verifiers$ and $\coordinator$ can compute a valid $\sigaggpr$. Alice eventually outputs $tx_A$. As described in Case 4, Alice can only pass smart contract verification if she outputs a valid $tx_A$. $\env$'s view will be $(l, \sigma_{A}, h, pk_{A}, \pi, y, tx_{A}, block_{curr}$, $pp$, $t_1, t_2, \alpha$, $\sigma_{V_i}$, $\messages_{i}$, $\sigma_{V^{\prime}_{i}}$, $M_{i}^{\prime})$ for $i \in [1 \dots | \verifiers | ]$.

    b) \textbf{Ideal-world}: $\simu$ needs to simulate the actions of $\verifiers$ to $\env$. $\env$ picks a $M_A$ and sends to $\simu$, the hash $h$ corresponding to $M_A$, $\sigma_A$, and $l$. If $h$ or $\sigma_A$ are invalid then $\simu$ would not generate $\verifiers$ signatures for $\env$. $\env$ can decide to proceed with the protocol by generating a corrupt $\sigagg$ but this would fail verification in the $code$ of the $SC.id$ smart contract. If valid request and $l$ is received from $\env$, $\simu$ sends $\verifiers$ signatures to $\env$ and $\env$ can generate $\sigagg$. If $\env$ decides to send corrupt proof to $\simu$, then it would fail verification and $\simu$ would not generate corresponding accept signatures from $\verifiers$ to send to $\env$. Like the previous stage, the $\env$ can create corrupt $\sigaggpr$ but this would fail verification eventually at the $SC.id$ smart contract.
    The only way for $\env$ to proceed is to compute valid $\VDF$ proof. Upon receiving valid proof $\simu$ verifies it and generates $\verifier$ signatures which are forwarded to $\env$. $\env$ generates a valid $\sigaggpr$, creates $tx_A$, and submits to $\F{bc}$.  As described in Case 4, $\env$ can only pass smart contract verification if  $tx_A$ contains valid signatures and $M_A$. $\env$'s view will be $(l, \sigma_{A}, h, pk_{A}, p, s, tx_{A}, block_{curr}$, $pp$, $t_1, t_2, \alpha$, $\sigma_{V_i}$,  $\messages_{i}$, $\sigma_{V^{\prime}_{i}}$, $M_{i}^{\prime})$ for $i \in [1 \dots | \verifiers | ]$.
    
        
        \noindent
        \textbf{8) Case 7: Alice, $\coordinator$ and some verifiers are corrupt}
   
    a) \textbf{Real-world}:  If Alice sends corrupt $\sigma_A$ to the $\coordinator$ and $\verifiers$, or if Alice sends a valid request but $\coordinator$ chose a corrupt $l$ similar to Case 2, the honest majority $\verifiers - \verifpr$ will not send accept signatures to $\coordinator$ since the request or $l$ will not pass verification. The $\coordinator$ can create corrupt $\sigagg$ but this would fail verification eventually at the smart contract. If the $\sigma_A$ is valid and the $\coordinator$ picks valid $l$, the honest majority $\verifiers - \verifpr$ will reply with accept messages so $\coordinator$ can generate a valid $\sigagg$. 
    Alice on receiving the $\sigagg$ can choose to send an invalid $\VDF$ proof which would not generate accept signatures from the honest majority $\verifiers - \verifpr$. Like the previous stage, the $\coordinator$ can create corrupt $\sigaggpr$ but this would fail verification eventually at the smart contract.
    If Alice computed a valid $\VDF$ proof, the $\coordinator$ would receive accept signatures from the honest majority $\verifiers - \verifpr$ and $\coordinator$ can compute a valid $\sigaggpr$. Alice eventually outputs $tx_A$. As described in Case 4, Alice can only pass smart contract verification if she outputs a valid $tx_A$. $\env$'s view will be $(l, \sigma_{A}, h, pk_{A}, \pi, y, tx_{A}, block_{curr}$, $pp$, $t_1, t_2, \alpha$,$\sigma_{V_i}$, $\messages_{i}$, $\sigma_{V^{\prime}_{i}}$, $M_{i}^{\prime})$ for $i \in [1 \dots | \verifiers | ]$.
    
    b) \textbf{Ideal-world}:
    $\simu$ needs to simulate the actions of honest majority $\verifiers - \verifpr$ to $\env$. As in Case 6, $\simu$ will not send any accept messages to $\env$ when simulating the honest verifiers if $\env$ sends corrupt $\sigma_A$ and $h$ on behalf of Alice and corrupt $l$ on behalf of the $\coordinator$.
    $\env$ will not be able to create valid $\sigagg$ and $\sigaggpr$ since the honest majority of verifiers will be simulated by $\simu$. 
    The only way for $\env$ to proceed is to submit valid $h, \sigma_A$ and $l$ to $\simu$ and compute a valid $\sigagg$ and $\VDF$ proof. Upon receiving valid proof $\simu$ verifies it and generates $\verifiers - \verifpr$ signatures which are forwarded to $\env$. $\env$ generates a valid $\sigaggpr$, creates $tx_A$, and submits to $\F{bc}$.  As discussed in Case 4, $\env$ can output a corrupt $tx_A$ but the verification checks in the smart contract $code$ will fail. $\env$'s view will be $(l, \sigma_{A}, h, pk_{A}, p, s, tx_{A}, block_{curr}$, $pp$, $t_1, t_2, \alpha$, $\sigma_{V_i}$,  $\messages_{i}$, $\sigma_{V^{\prime}_{i}}$, $M_{i}^{\prime})$ for $i \in [1 \dots | \verifiers | ]$.

Let $\mathbb{TX}_{\adv}$ represent the set of $\adv$'s transactions, $\mathbb{TX}_{\env}$ represent the set of $\env$'s transactions, $\mathbb{TX}_{\simu}$ represent the set of $\simu$'s transactions, and let $tx_{A}$ represent Alice's transaction in the pending pool, i.e., $\txpooltable$, respectively. Let the time elapsed since $tx_A$ entered the $\txpooltable$ be denoted by $T_E$, the maximum miner fee allowed per block by the blockchain system be $\bcblockmaxfee$, let $t_1$ be the time taken for computing a given VDF, and let $\mathcal{P}(\mathbb{A})$ denote the power set of set $\mathbb{A}$. Then the advantage of $\adv$ in winning the $\first{}$ game against Alice, i.e., Alice's $tx_{A}$ getting frontrunned is given by the following inequality:

\begin{equation*}
\label{eqn:advantage}
\begin{split}
\advantage{}{\adv,\first{}}[(\lambda)] & \leq Pr \Big[ 
\Big( \mathbb{TX}_B =  min(|\mathbb{X}|); \\
& \mathbb{X} \in \mathcal{P}((\mathbb{TX}_\env \cup \mathbb{TX}_\adv ) \cup (\mathbb{TX}_\simu \setminus \{tx_a\}))\Big)  \wedge \\
& \Big( \forall ~tx \in \mathbb{TX}_B,  ~\sum tx.txfee~ \lessapprox \bcblockmaxfee \Big) \\
& \wedge (T_E >t_1)
\Big ].
\end{split}
\end{equation*}

\end{document}